(9 postcript files come as part 2, (fig1, fig3..10), uuencoded)

\documentstyle[12pt]{article}

\newcommand{\Sv}{{\bf S}}
\newcommand{\R}{{\bf R}}
\newcommand{\A}{{\cal A}}
\newcommand{\acos}{\mbox{acos}}

\newcommand{\bee}{\begin{equation}}
\newcommand{\ee}{\end{equation}}
\newcommand{\beea}{\begin{eqnarray}}
\newcommand{\eea}{\end{eqnarray}}
\newcommand{\rme}{{\rm e}}
\begin{document}
\begin{titlepage}
\thispagestyle{empty}
\parskip=12pt
\raggedbottom

\def\mytoday#1{{ } \ifcase\month \or
 January\or February\or March\or April\or May\or June\or
 July\or August\or September\or October\or November\or December\fi
 \space \number\year}
\noindent
\hspace*{11cm} BUTP--93/17\

\vspace*{1cm}


\begin{center}

{\LARGE Perfect lattice action  \\
\vspace*{0.5cm}
for asymptotically free
theories}\footnote{Work supported in part by Schweizerischer
Nationalfonds}

\vspace{2cm}

P. Hasenfratz and F. Niedermayer\footnote{on leave from the
Institute  for Theoretical Physics,
E\"otv\"os University, Budapest}\\

\vspace{1cm}

Institute for Theoretical Physics \\
University of Bern \\
Sidlerstrasse 5, CH-3012 Bern, Switzerland

\vspace{1cm}

July 1993
\\ \vspace*{1cm}

\nopagebreak[4]

\begin{abstract} \normalsize

There exist lattice actions which give cut--off independent
physical predictions even on coarse grained lattices. Rotation
symmetry is restored, the spectrum becomes exact and, in
addition, the classical equations have scale invariant
instanton solutions. This perfect action can be made short
ranged. It can be determined by combining analytical
calculations with numerical simulations on small lattices.
We illustrate the method and the benefits on the $d=2$
non--linear $\sigma$--model.

\end{abstract}
\end{center}
\end{titlepage}

\section{Introduction and Summary}

In many theoretical problems in physics, at a given stage,
numerical analysis is necessary. Numerical problems require some
regularization, most often different meshes are used.
If the lattice is coarse, the numerical procedure is relatively
easy, but the results are contaminated by the artefacts of the
regularization. It is a difficult problem to remove these
artefacts by making the meshes systematically finer and then
extrapolating.

We shall discuss the problem of artefacts in the context of
lattice regularized asymptotically free quantum field theories.
Removing the cut--off effects to reach the proper continuum
limit is a central problem in this field \cite{1}.
The method we describe, however, might find applications
elsewhere, for example in solving partial differential
equations.

In the study of quantum field theories cut--off effects show up
in different ways. If the lattice is coarse grained then
all the results are distorted, independently whether short or
long distances are involved. Similarly, even on a fine lattice,
at distances of a few lattice units the lattice structure,
rather than physics dominates the results.

Removing the cut--off effects requires a careful analysis.
Fig.~1 illustrates this process on the example of the running
coupling in the $d=2$ non--linear $\sigma$--model \cite{2}.
The renormalized coupling at the distance
scale $L$ is identified here with
$m(L)L$, where $m(L)$ is the mass gap in a periodic box of
size $L$. For small $L$ ($L \ll 1/m_{\infty}$, where $m_{\infty}$
is the mass in the infinite volume)
this quantity can be studied in
perturbation theory and shows the properties of an asymptotically
free coupling: as $L$ is changed, the coupling runs according
to the perturbatively known beta--function. For larger values
of $L$ the scale dependence of the coupling, for example the
relation between $g(2L) \equiv m(2L)\cdot 2L$ and
$g(L) \equiv m(L)\cdot L$, requires a numerical calculation.
A lattice is introduced with lattice unit $a$, and for
a given $g(L)$ (in Fig.~1 $g(L)=1.0595$) $g(2L)$ is calculated
on lattices with $L/a=5,6,7,\ldots,16$ and the results
are extrapolated to $L/a=\infty$. This is a non--trivial
numerical problem even in $d=2$.

A possibility to reduce lattice artefacts is to use improved
actions. The idea behind this procedure is that for a large
class of actions the physical predictions are universal
while the cut--off effects are not.
Although the history of improved actions is long \cite{3a,3b}
they had a limited effect on actual calculations until now.
The reason is that in some cases the proposed actions were
derived using ad hoc assumptions,
or uncontrolled approximations, in other cases  they were
fixed by perturbation theory and it is not clear whether in
typical simulations they improve and to what extent.
Deep in the continuum limit a perturbatively fixed action certainly
improves. This is, however, not necessarily the case at moderate
correlation lengths.
Fig.~1 illustrates this problem also. Perturbation theory fails
to reproduce even the sign of the observed cut--off effects \cite{2}.

A radical solution would be to use a {\it perfect} lattice
action which is completely free of lattice artefacts.
That such perfect actions exist follows from Wilson's
renormalization group  (RG) theory \cite{4,5}.
The aim of this paper is to show that in asymptotically free
theories a combination of analytical and numerical techniques
allows finding the perfect action to a sufficient precision.
We consider an action perfect in practical calculations if
even on a coarse grained lattice (correlation length
=O($a$)), no cut--off effects can be seen in numerical simulations.
The perfect action is not unique. This fact can be used to find
a perfect action whose range of interaction is short
and whose structure is relatively simple. Simulating such actions
the gain/cost ratio can be very large.

In this work we study this problem in the $d=2$, O(3) non--linear
$\sigma$--model, which we consider a pilot project, a preparation
for $d=4$ Yang--Mills theories and QCD.
A quadratic lattice is considered and the partition function
is written as
\bee
Z=\prod_n \int d\Sv_n \delta (\Sv_n^2 -1)\rme^{-\beta\A(\Sv)}\,,
\label{1}
\ee
where $\beta\A(\Sv)$ is some lattice representation of the
continuum action
\bee
\beta\A^{cont}(\Sv)={\beta\over 2} \int d^2x
\partial_{\mu} \Sv(x) \partial_{\mu} \Sv(x) \,.
\label{2}
\ee
Beyond the basic requirements of O(3) symmetry, locality,
correct classical limit, translation and $90^\circ$--rotation
symmetry, the form of the lattice action is largely arbitrary.
It might contain nearest neighbour, next--to--nearest
neighbour, etc., even different multispin interactions.
Let us denote the corresponding couplings by
$c_1,~c_2,\ldots$. The action $\beta\A(\Sv)$ is represented
by a point in the infinite dimensional space of couplings
($\beta,c_1,c_2,\ldots$).
We shall consider RG transformations in
configuration space, namely block transformations with a scale
factor of 2. Under repeated block transformations the action
moves in this coupling constant space.
The expected flow diagram is sketched in Fig.~2 \cite{5,6}.
In the $\beta=\infty$ hyperplane there is a fixed point (FP)
$c_1^*,c_2^*,\ldots$, whose exact position depends on the details
of the block transformation. We shall use the notation
$\A(\Sv;c_1^*,c_2^*,\ldots)=\A^*(\Sv)$ and call
$\beta \A^*(\Sv)$ the FP--action. The FP has one marginal and
infinitely many irrelevant directions. The marginal operator
is $\A^*$ itself \cite{5}. Actually, $\A^*$ is not exactly marginal,
it is weakly relevant. If $\beta$ is very large, under a
RG transformation with a scale factor 2,
$\beta\A^*(\Sv) \to \beta'\A^*(\Sv)$, where
$\beta'=\beta-{1\over 2\pi}\ln 2$.
The coupling $g=1/\beta$ grows according to the beta--function
of this asymptotically free theory. The trajectory which leaves
the FP along the weakly relevant direction is called the
renormalized trajectory (RT). For large $\beta$ the RT runs
along the FP--action, but for smaller $\beta$ they do not coincide
anymore.

\setcounter{figure}{1}
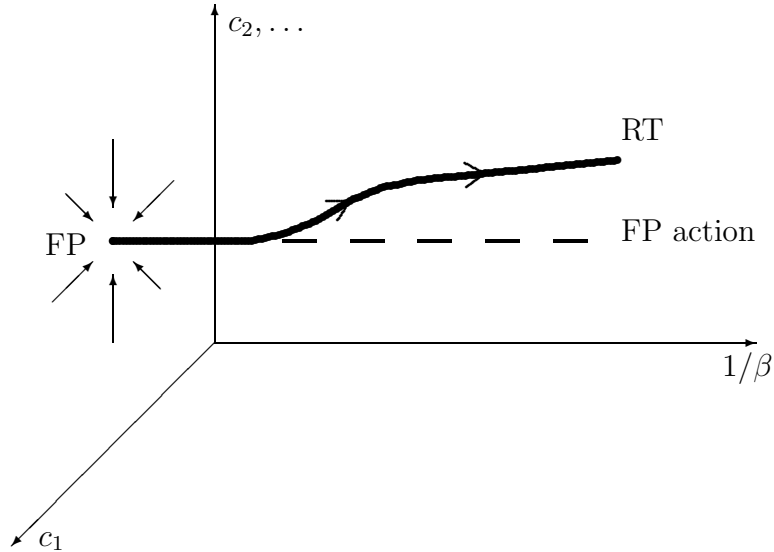
\begin{figure}  
\setlength{\unitlength}{.9mm}
\begin{center}
\begin{picture}(120,80)
\put(30,30){\vector(1,0){80}}
\put(30,30){\vector(0,1){50}}
\put(30,30){\vector(-1,-1){30}}
\put(4,0){$c_1$}
\put(32,76){$c_2,\ldots$}
\put(105,25){$1/\beta$}
\put(5,43){FP}
\put(15,45){\circle*{1}}
\put(15,30){\vector(0,1){10}}
\put(15,60){\vector(0,-1){10}}
\put(6,36){\vector(1,1){6}}
\put(24,54){\vector(-1,-1){6}}
\put(22,38){\vector(-1,1){4}}
\put(8,52){\vector(1,-1){4}}
\multiput(15,45)(.5,0){40}{\circle*{1}}
\multiput(35,45)(.5,.1){10}{\circle*{1}}
\multiput(40,46)(.5,.2){10}{\circle*{1}}
\multiput(45,48)(.5,.3){10}{\circle*{1}}
\multiput(50,51)(.5,.2){10}{\circle*{1}}
\multiput(55,53)(.5,.1){10}{\circle*{1}}
\multiput(60,54)(.5,0.05){60}{\circle*{1}}
\thicklines
\multiput(40,45)(10,0){5}{\line(1,0){5}}
\thinlines
\put(90,45){FP action}
\put(90,60){RT}
\multiput(50,51)(-.25,0){15}{\circle*{.5}}
\multiput(50,51)(-.13,-.21){15}{\circle*{.5}}
\multiput(70,55)(-.22,0.11){15}{\circle*{.5}}
\multiput(70,55)(-.20,-.14){15}{\circle*{.5}}
\end{picture}
\end{center}
\caption{Flow of the couplings under RG transformation
 in the O($N$) non--linear $\sigma$--model}
\end{figure}

It is easy to see that the points of the RT define perfect
actions. The argument goes as follows.
At any given $\beta$, the point of the RT is connected to
the infinitesimal neighbourhood of the FP by (infinitely many
steps of) exact RG transformations.
Since each step increases the lattice unit by a factor of 2,
{\it any} distance at the given $\beta$ (even 1 lattice unit)
corresponds to a long distance close to the FP.
The infinitesimal neighbourhood of the FP is in the
continuum limit, there are no cut--off effects there at long
distances. On the other hand, for all the questions which can
be formulated in terms of the degrees of freedom after the
transformation we get the same answer as before it.
So, there are  no lattice artefacts at the given $\beta$ on
the RT at any distances.

Since a RG step is a non--critical problem, the FP--action
and the actions on the RT in general are expected to be local
\cite{4,5}.
Locality, however, allows non--negligible interactions over
several lattice units in
the action. For practical reasons we neeed more than locality.
For practical applicability we have to answer positively
the following questions:
i.   Is it possible to determine $\A^*(\Sv)$ to a good precision?
ii.  Is $\A^*(\Sv)$ of sufficiently short ranged? Is the structure
of $\A^*$ simple enough allowing a parametrization where the
number of couplings remains relatively low, O(10--100)?
iii. Questions i. and ii. for the points of the RT.

Our pilot study shows that for the non--linear $\sigma$--model,
using a properly
chosen RG transformation, the answer to these questions is
'yes', even if we go down to small correlation lengths . The
determination of $\A^*(\Sv)$ is a saddle point problem which
requires minimalization over classical fields. The FP,
$\A^*(\Sv)$ is the perfect classical theory on the lattice:
it has stable instanton solutions whose action is independent
of the size $\rho$ of the instanton (scale invariance).
This should be the same in $d=4$ gauge theories also. This fact
might open new ways in the study of topological effects on the
lattice. The range of interaction in $\A^*(\Sv)$ depends on the
RG transformation. By a proper choice of the block--transformation
$\A^*(\Sv)$ becomes surprisingly short ranged. $\A^*(\Sv)$ contains
multispin couplings also but its structure is relatively simple.
 With O(20) couplings an excellent parametrization can be
obtained which works even on coarse configurations, i.e. at
small correlation lengths. The problems
related to $\A^*(\Sv)$ can be solved partly analytically,
which is a
special bonus in asymptotically free theories.

If $\beta$ is not very large, the FP--action
$\A_{FP}=\beta_{FP}\A^*(\Sv)$
is not a perfect action. It is, however, the {\it perfect classical}
action and
we expect that the lattice artefacts are significantly reduced.
Actually, we were not able to identify any cut--off effects when
simulating $\A_{FP}$ in our pilot study. The lattice artefacts in
the problem of the running coupling in Fig.~1 can be used,
for example, to illustrate how well the FP--action
performs. As Fig.~3 shows, even the coarsest discretization
$L/a=5$ gives a result which is consistent with the final
extrapolated value. Let us emphasize at this point that $\A_{FP}$
has nothing to do with perturbation theory.

The FP--action is 'almost perfect', which suggests that the RT
runs close to the FP--action even at moderate $\beta$--values.
In this case the blocked  FP--action will converge rapidly towards
the RT. We illustrate on an explicit numerical experiment that
the problem to find and parametrize the action after
a block--transformation at finite $\beta$ is feasible.
In traditional Monte Carlo simulations one goes deep in the
continuum limit, simulates large lattices and fights with memory
problems and critical slowing down. Our message is: do careful
calculations on small lattices instead.

There remained many interesting problems even in this $d=2$ model
which  we did not touch: an explicit study of the classical
solutions, the possibility of using strong coupling expansion
rather than Monte Carlo simulations for the perfect action at
small correlation lengths and, most notably, the problem of
constructing currents corresponding to the perfect action.

The idea of constructing improved actions using RG ideas is
old \cite{3b,5}.
But, perhaps because the problem was considered as not
really feasible, already at the start crude approximations and
assumptions were introduced. We would like to show here that
--- due to the specific properties of asymptotically
free theories --- one can get close to the dream of a perfect
regularization. First, finding the FP is
a relatively easy {\it classical} problem.
This is a significant advantage over other statistical
models, e.g. the $d=2$ Ising model \cite{SG}. Second, even a
simple optimalization makes the FP--action short ranged. Third,
the FP--action, which is the perfect classical action, performs
very well even at small correlation lengths. This feature makes
it a good first approximation in finding the RT.

Earlier RG and Monte Carlo RG studies \cite{7,8} revealed many
analogies between $d=4$ Yang--Mills theories and $d=2$
asymptotically free spin models. This gives a hope that the
results of this paper can be generalized. Of course, the final
goal is
to treat full QCD where the cut--off effects are even more
troublesome. The formalism can be extended to fermions.
The free fermion problem has been discussed in \cite{9},
but the real hard work is to be done.

We should mention that the physical picture in the
multigrid--RG program as formulated by Mack \cite{Ma} has
close analogies to that discussed here.


\section{The fixed point and the fixed point action}

\subsection{The  equation for $\A^*$}

We shall define the RG transformation in the model
in eq.~(\ref{1}) as follows.
The square lattice is divided into $2 \times 2$ blocks and
to every block we associate a block spin $\R_{n_B}$,
which is a certain average of the four original spins in the
block. We shall denote the points of the original
and the block lattice by $n$ and $n_B$, respectively.
The spins $\Sv_n$ and $\R_{n_B}$ are
normalized O$(N)$
vectors. The RG transformation is defined as
\bee
\rme^{-\beta'\A'(\R)}=\int_{\Sv}
\exp \left\{ -\beta\A(\Sv) +\sum_{n_B}
\left[ P\R_{n_B}\sum_{n\in n_B}\Sv_n
-\ln Y_N(P|\sum_{n\in n_B}\Sv_n|) \right] \right\} \,,
\label{3}
\ee

$$
\int_{\Sv}=
\prod_n \int d\Sv_n \delta\left(\Sv_n^2-1\right)\,,
$$
where $P$ is a parameter of the transformation, while the
normalization factor $Y_N(z)$ assures that the partition
function remains unchanged up to an irrelevant constant:
\bee
\int d\R \delta\left( \R^2-1\right) \rme^{\R {\bf b}}
=\mbox{const} \cdot  Y_N\left( |{\bf b}| \right) \,.
\label{4}
\ee

The function $Y_N(z)$ is related to the modified Bessel
function (some of its properties are summarized in
Appendix F in \cite{10}, for example), specifically
$Y_3(z) \sim$ sinh(z)/z. For $P \rightarrow \infty$,
the block--transformation goes over to a
$\delta$--function giving
$\R_{n_B}=\sum_{n\in n_B}\Sv_n/|\sum_{n\in n_B}\Sv_n|$.
We keep, however, $P$ finite and write
\bee
P=\beta\cdot\left[ \kappa+{\rm O}\left(1/\beta\right)\right]
\,,
\label{5}
\ee
where $\kappa$ is a free parameter. Similar RG transformations
were considered before in Monte Carlo RG studies \cite{11}.
The coupling $\beta$ is defined by the
requirement that on very smooth configurations the action goes
over to the form in eq.~(\ref{2}) (see also section 2.4).

In the limit $\beta \rightarrow \infty$, eq.~(\ref{3}) can be
written as
\bee
\rme^{-\beta'\A'(\R)}=\int_{\Sv}
\exp\left( {-\beta \left\{ \A(\Sv)-\kappa \sum_{n_B}
\left[ \R_{n_B}\sum_{n\in n_B}\Sv_n
- | \sum_{n\in n_B}\Sv_n | \right] \right\} }\right) \,,
\label{6}
\ee
where $\beta'=\beta-\mbox{O}(1)$, due to asymptotic freedom.
Eq.~(\ref{6}) is a saddle point
problem in this limit, giving
\bee
\A'(\R)=\min_{\{\Sv\}}
\left\{ \A(\Sv)-\kappa \sum_{n_B}
\left[ \R_{n_B}\sum_{n\in n_B}\Sv_n
- | \sum_{n\in n_B}\Sv_n | \right] \right\} \,.
\label{7}
\ee
The FP of the transformation is determined by the equation
\bee
\A^*(\R)=\min_{\{\Sv\}}
\left\{ \A^*(\Sv)-\kappa \sum_{n_B}
\left[ \R_{n_B}\sum_{n\in n_B}\Sv_n
- | \sum_{n\in n_B}\Sv_n | \right] \right\} \,.
\label{8}
\ee
As we discussed at length in the Introduction, the FP--action
$\beta\A^*(\R)$ -- although it is not 'perfect' at finite
$\beta$ -- is an excellent starting action for further studies
even at moderate $\beta$--values
(at moderate correlation lengths).
So, we shall need $\A^*(\R)$ for strongly fluctuating
configurations also. Observe that eq.~(\ref{6}) reduces to the
saddle point equation eq.~(\ref{7}) for any configuration ${\R}$.
If the configuration ${\R}$ is strongly
fluctuating, then the minimizing configuration
 ${\Sv}$ will not be smooth either.
In general, eqs.~(\ref{7}),(\ref{8})
and their solutions have nothing to do with
perturbation theory.

Eq.~(\ref{8}) for $\A^*$ is a highly non--trivial equation.
Some of the important properties of the solution can be
obtained, however, without solving the equation explicitly.

\subsection{$\A^*$ as a perfect classical theory on the lattice}

Using eq.~(\ref{8}) it is easy to show the
following\footnote{We are indebted to Uwe Wiese who raised
our attention to the classical solutions in this context.}

\noindent{\bf Statement}\\
If the configuration $\{\R\}$ satisfies the FP classical
equations (i.e. the classical equations corresponding to
$\A^*$) and it is a local minimum of $\A^*(\R)$
(allowing also for zero modes)
then the configuration $\{\Sv(\R)\}$  on the finer
lattice which minimizes the right hand side of
eq.~(\ref{8}) satisfies the FP equations as well.
In addition, the value of the action remains unchanged,
$\A^*(\Sv(\R))=\A^*(\R)$.

\noindent{\bf Proof}\\
Since $\{\R\}$ is a solution of the classical equations
of motion, $\delta\A^*/ \delta\R = 0$,
the configuration $\{\Sv\}=\{\Sv(\R)\}$
should satisfy the equation
\bee
\sum_{n\in n_B} \Sv_n = \lambda_{n_B}\R_{n_B} \,,
\label{9}
\ee
for any $n_B$. Here $\lambda_{n_B}\geq 0$ since we excluded
the case when $\A^*(\R)$ is a local maximum in $\R_{n_B}$.

Since the expression
\bee
\kappa \sum_{n_B}\left[ |\sum_{n\in n_B} \Sv_n|
-\R_{n_B}\sum_{n\in n_B} \Sv_n \right]
\label{12}
\ee
in eq.~(\ref{8}) takes its absolute minimum (zero) on the
configuration $\{\Sv(\R)\}$ satisfying eq.~(\ref{9}), $\{\Sv(\R)\}$
is also a stationary point of $\A^*$,
\bee
\left. {\delta\A^*(\Sv)\over
\delta\Sv} \right|_{\Sv=\Sv(\R)}=0\,,
\label{13}
\ee
with the same value of the action,
$\A^*(\Sv(\R))=\A^*(\R)$, what we wanted to show.

According to this Statement, to any solution $\{\R\}$ (which is
a local minimum)
with a characteristic size $\rho$, there exists another
solution $\{\Sv(\R)\}$ of size $2\rho$ with exactly
the same value of the action $\A^*$. It is natural to assume
that this solution on the finer lattice is also a local
minimum.
Specifically, for $N=3$ this observation implies that if
$\A^*$ has an instanton solution of size $\rho$ then
there exist instanton solutions of size
$2\rho,2^2\rho,\ldots,2^k\rho,\ldots$ with the value of
the action being exactly $4\pi$ for all these instantons.
The value $4\pi$ follows from the fact
that very large instantons are smooth on the lattice
and then any valid lattice representation of eq.~(\ref{2})
gives the continuum value.

The standard action has classical solutions of antiferromagnetic
character --- e.g. when one of the spins points downwards in a
background of upward pointing spins --- which are smooth local
maxima in some of the spins. We expect, although we are not
able to prove, that $\A^*$ does not have such solutions.
For example, the configuration mentioned above is not
a solution since $\delta\A^*(\R)/\delta\R$ does not vanish at
$\vartheta=\pi$, as shown in Fig.~6.

It was observed earlier in \cite{12} that using an approximate RG
improved action the stability of instantons can be increased
significantly. This result is in accordance with the Statement
above.

It is important to observe that the reverse of the Statement
is {\it not} true:
if the configuration $\{\Sv_n\}$ is a solution, then the
configuration $\{\R_{n_B}\}$, where
$\R_{n_B}=\sum_{n\in n_B} \Sv_n / |\sum_{n\in n_B} \Sv_n|$,
is not necessarily a solution. The proof fails because
for this configuration $\{\R_{n_B}\}$ the minimizing
configuration $\{\Sv(\R)\}$ in eq.~(\ref{8}) is not
necessarily equal to $\{\Sv_n\}$ itself. One can show
that $\{\Sv_n\}$ is a minimum but not always the absolute
minimum. Actually, this is the mechanism which prevents the
existence of arbitrarily small instanton solutions
on the lattice \cite{13}.

With respect to the classical solutions, $\A^*$ has the
same scale symmetry as the continuum action. We shall
give later further arguments which show that it is
a perfect classical lattice representation of the
continuum action.

\subsection{Parametrization}
Eq.~(\ref{8}) defines the value of the FP, $\A^*$ for any
configuration ${\R}$ and, as we shall discuss later,
one can advise numerical
procedures which give this value to high precision.
However, if we want to use the FP action $\beta_{FP} \A^*$
in numerical simulations we can not avoid to parametrize
$\A^*$ by a (limited) number of coupling constants.
We can write in general
\beea
\A^*(\Sv) & = & -{1\over 2} \sum_{n,r} \rho(r)
\left( 1-\Sv_n \Sv_{n+r}\right)
\label{14} \\
& & + \sum_{n_1,n_2,n_3,n_4} \!\!\!c(n_1,n_2,n_3,n_4)
\left( 1-\Sv_{n_1} \Sv_{n_2}\right)
\left( 1-\Sv_{n_3} \Sv_{n_4}\right) +\ldots \nonumber
\eea
where the summations go over all the lattice points. It is a
significant help in the parametrization and optimalization
problem that the first two functions $\rho$ and $c$ in
eq.~(\ref{14}) can be calculated directly. We shall discuss
this problem first.

\subsection{The determination of $\rho(r)$}
Consider a configuration ${\R}$ where the spins fluctuate
around the first axes:
\bee
\R_{n_B}=\left( {\sqrt{1-\vec{\chi}_{n_B}^2}
\atop \vec{\chi}_{n_B}} \right)
\label{15}
\ee
where $\vec{\chi}_{n_B}$ has $(N-1)$ components, and
$|\vec{\chi}_{n_B}|\ll 1$. In this case, the saddle--point
configuration in eq.~(\ref{8}) will also fluctuate around
the first axes, so we can write
\bee
\Sv_{n}=\left( {\sqrt{1-\vec{\pi}_{n}^2}
\atop \vec{\pi}_{n}} \right)
\label{16}
\ee
where $|\vec{\pi}_{n}|\ll 1$. The terms quadratic in
$\vec{\pi}$ and $\vec{\chi}$ give a closed equation for
$\rho$ when eqs.~(\ref{8})  and (\ref{14}) are used:
\bee
{1\over2}\sum_{n_B,r_B} \rho(r_B)
\vec{\chi}_{n_B}\vec{\chi}_{n_B+r_B}=
\min_{ \{ \vec{\pi}\} } \left\{
{1\over2}\sum_{n,r} \rho(r)\vec{\pi}_n\vec{\pi}_{n+r}
+2\kappa\sum_{n_B}\left(\vec{\chi}_{n_B}-
{1\over 4}\sum_{n\in n_B} \vec{\pi}_n \right)^2
\right\} \,.
\label{17}
\ee
The equation for $\rho$ is independent of $N$ as far
as  $N>1$.
Eq.~(\ref{17}) is just the FP equation of a free scalar
theory with a Gaussian block--transformation:
\beea
& & c\cdot\exp \left\{
-{1\over2}\sum_{n_B,r_B} \rho(r_B)
\vec{\chi}_{n_B}\vec{\chi}_{n_B+r_B}
\right\}
= \label{18} \\
& & \prod_n\int d\vec{\pi}_n
\exp \left\{-2\kappa\sum_{n_B}\left(\vec{\chi}_{n_B}
-{1\over 4}\sum_{n\in n_B} \vec{\pi}_n \right)^2
-{1\over2}\sum_{n,r} \rho(r)\vec{\pi}_n\vec{\pi}_{n+r}
\right\} \,.
\nonumber
\eea
Eqs.~(\ref{17}) and (\ref{18}) are equivalent: Gaussian
integrals can be replaced by minimalizations. The free
field problem has been studied and solved in \cite{14} a long
time ago. In Fourier space the solution has the form
\bee
{1\over \tilde{\rho}(q)}=
\sum_{l = -\infty}^{+\infty}
{1\over (q +2\pi l)^2}
\prod_{i=0}^1 {\sin^2(q_i/2)\over (q_i/2+\pi l_i)^2}
+{1\over 3\kappa} \,,
\label{19}
\ee
where the summation is over the integer vector $l=(l_0,l_1)$,
$(q +2\pi l)^2=(q_0 +2\pi l_0 )^2+(q_1 +2\pi l_1 )^2$ and
\bee
\rho(n)=\int_{-\pi}^{\pi}{d^2q\over (2\pi)^2}
\rme^{iqn} \tilde{\rho}(q) \,.
\label{20}
\ee

The normalization of $\rho$ is fixed by demanding
$\tilde{\rho}(q) \rightarrow q^2$ for small $q$.
In configuration space this corresponds to
$\sum_r \rho(r) r^2=-4$. We fix the value of
$\rho(0)$ by the convention $\sum_r \rho(r)=0$.

There is an elegant, simple way to derive eq.~(\ref{19})
which we discuss below, since it might simplify the
analogous problem for gauge and fermion fields also.
A similar trick was used in \cite{15} in an approximate RG
calculation in the large--$N$ limit with
$\delta$--function blocking.

The observation is that in a free field theory the
two--point function is directly related to the action.
On the other hand, the two--point function at the FP
can be calculated easily. From eq.~(\ref{18}) it follows
\bee
\langle \chi_{n_B}\chi_{n_B'} \rangle
=\left({1\over 4}\right)^2
\sum_{n\in n_B \atop n'\in n_B'}
\langle \pi_{n}\pi_{n'} \rangle +
{1\over 4\kappa} \delta_{n_B n_B'} \,.
\label{21}
\ee
Using this relation recursively, after $j$ RG steps we get
\bee
\langle \chi_{n_B}\chi_{n_B'} \rangle
=\left({1\over 4}\right)^{2j}
\sum_{n\in n_B \atop n'\in n_B'}
\langle \pi_{n}\pi_{n'} \rangle +
{1\over 4\kappa}\left(1+\frac{1}{4}+\frac{1}{4^2}
+\ldots + \frac{1}{4^j}\right)
 \delta_{n_B n_B'} \,.
\label{22}
\ee
where $\chi$ is the block--field after $j$ steps, and the
block $n_B$ contains $4^j$ original lattice points.
Taking the $j \rightarrow \infty$ limit, and considering
the original lattice infinitely fine, the summations in
the first term on the right--hand side of eq.~(\ref{22})
go over to integrals. The integrals are over a block
$(a \times a)$ centered around the points $n_B$ and $n_B'$,
respectively. For the two--point function on the infinitely
fine lattice we can use the standard $1/k^2$ propagator.
Taking $a=1$, and simplifying the notations we get
\bee
\langle \chi_{n}\chi_{n'} \rangle
=\int_{-1/2}^{1/2} d^2x
\int_{-1/2}^{1/2} d^2x'
\int_{-\infty}^{\infty}
\frac{d^2k}{(2\pi)^2}
\frac{\rme^{ik(n+x-n'-x')}}{k^2}
+\frac{1}{3\kappa}
 \delta_{n n'} \,.
\label{23}
\ee
After many RG steps the system runs to the FP (taking the
mass to be zero at the beginning, only irrelevant
directions remain), therefore
\bee
\langle \chi_{n}\chi_{n'} \rangle
=\int_{-\pi}^{\pi}
\frac{d^2q}{(2\pi)^2}
\frac{\rme^{q(n-n')}}{\tilde{\rho}(q)}
 \,.
\label{24}
\ee

Dividing the $k$--integration into an integration and
summation by introducing the integer vector
$l=(l_0,l_1)$
\bee
k_i=q_i+2\pi l_i,~~~~q_i\in (-\pi,\pi)\,, ~~~~i=0,1\,,
\label{25}
\ee
and performing the integrals over $x$ and $x'$, the result
in eq.~(\ref{19}) follows. Using similar steps, the
two--point function in configuration space has been
obtained for $\delta$--function blocking by Iwasaki \cite{3b}
and by Gawedzki and Kupiainen \cite{15a} also.

Let us remark that the $1/4$ factor in the block
transformation of a free theory in eq.~(\ref{18})
is replaced by $b=2^{-(d+2)/2}$ in $d$ dimensions.
Only this choice leads to a FP \cite{4,14}.
In the non-linear $\sigma$--model, the factor $1/4$
in the last term in eq.~(\ref{17}) is fixed.
That these two factors coincide for $d=2$ is related
to the fact that in two dimensions (and only in this case)
the non--linear $\sigma$--model is asymptotically free.

\subsection{The properties of $\rho(r)$; the perfect
Laplace operator on the lattice}

The properties (most notably the range of interaction) of
$\rho(r)$ depend on the free parameter $\kappa$. The choice
$\kappa \approx 2.$ is optimal. For $\kappa=2$, $\rho(r)$
decays very rapidly with growing $|r|$, like
$\sim \exp (-3.44|r|)$, Fig.~4. For this choice  $\rho(r)$
is strongly dominated by the nearest neighbour and diagonal
couplings while the couplings at distance $> 1$ are already
small. As Table~1 shows, $\rho(3,3)$ is, for example, $5$
orders of magnitude smaller than the nearest neighbour
coupling. The $\kappa=2$ value is distinguished also by
the fact that in $d=1$ (or, equivalently, in any $d$ on
configurations which depend only on one coordinate) this
choice leads to a simple nearest neighbour action.
Fig.~4 also shows that the choice $\kappa=\infty$
(corresponding to a block transformation with
$\delta$--function) gives a considerably larger interaction
range, $\rho(r)\sim \exp (-1.45|r|)$.

\begin{table}  
\begin{center}
\begin{tabular}{|c|c||c|c|}
\hline
$r$ & $\rho(r)$ & $r$ & $\rho(r)$ \\
\hline
(1,0) & $-0.61802 $           & (4,0) & $-2.632\cdot 10^{-6}$ \\
(1,1) & $-0.19033 $           & (4,1) & $ 7.064\cdot 10^{-7}$ \\
(2,0) & $-1.998\cdot 10^{-3}$ & (4,2) & $ 1.327\cdot 10^{-6}$ \\
(2,1) & $-6.793\cdot 10^{-4}$ & (4,3) & $-7.953\cdot 10^{-7}$ \\
(2,2) & $ 1.625\cdot 10^{-3}$ & (4,4) & $ 6.895\cdot 10^{-8}$ \\
(3,0) & $-1.173\cdot 10^{-4}$ & (5,0) & $-8.831\cdot 10^{-8}$ \\
(3,1) & $ 1.942\cdot 10^{-5}$ & (5,1) & $ 3.457\cdot 10^{-8}$ \\
(3,2) & $ 5.232\cdot 10^{-5}$ & (5,2) & $ 3.491\cdot 10^{-8}$ \\
(3,3) & $-1.226\cdot 10^{-5}$ & (5,3) & $-3.349\cdot 10^{-8}$ \\
      &                       & (5,4) & $ 8.408\cdot 10^{-9}$ \\
      &                       & (5,5) & $-1.657\cdot 10^{-10}$ \\
\hline
\end{tabular}
\end{center}
\caption{The couplings of the quadratic two--spin interaction
terms at a distance $r=(r_0,r_1)$ for the optimal choice of the
block transformation with $\kappa=2$.
Note that in our convention, for the standard action
the  only non--vanishing entry in this list would be
$\rho_{ST}(1,0)=-1$.}
\end{table}

A free field theory defined in terms of $\rho$ on the
lattice gives a cut--off independent account of the
physical properties of the theory. The spectrum is exact.
In the continuum the energy of an excitation with
momentum $k_1$ is $E(k_1)=|k_1|$, where
$k_1 \in(-\infty,\infty)$ (massless relativistic free
particle). Using eq.~(\ref{19}) one can check easily that
the lattice field theory with $\rho$ has exactly this
spectrum. (A possible method is to calculate the
two--point function at large time--separation.)
Although $q_1$ in eq.~(\ref{19}) lies in $(-\pi, \pi)$
only, the full spectrum is reproduced because for every
given $q_1$ there are infinitely many energy eigenstates.
This is related to the summation over $l$ in
eq.~(\ref{19}).

Fig.~5 compares the predictions for the energy spectrum
of $\rho$ (exact) to those of the standard lattice
regularization, Symanzik tree level improved action
and the approximation where the $\rho(1,0)$ and
$\rho(1,1)$ couplings are kept only.\footnote{We are
indebted to M. Blatter for his help on Fig.~5.}
Already this crudest approximation performs very well.

The finite difference operator
$\left. \Delta_L  \Phi \right)_n =
-\sum_r \rho(r) \Phi_{n+r}$
is a perfect discretization of the continuum Laplacian.
Let $\phi(x)$ be a field in the continuum satisfying
the Laplace equation
\bee
\Delta \phi(x)=0 \,.
\label{26}
\ee

Lay a lattice on the $d=2$ continuum plane with
lattice unit 1 and define on the lattice point $n$ a
lattice field
\bee
\Phi_n =
\int_{-\frac{1}{2}}^{\frac{1}{2}} d^2x
\phi(n+x)
\label{27}
\ee
The field $\Phi_n$ satisfies
\bee
\Delta_L \Phi =0 \,,
\label{28}
\ee
and gives the same action as $\phi(x)$:
\bee
-\int d^2x \phi \Delta \phi =
-\sum_n \Phi_n \left( \Delta_L \Phi \right)_n
\label{29}
\ee

\subsection{The determination of $c(n_1,n_2,n_3,n_4)$}

In order to obtain the coefficients $c(\ldots)$ we have
to expand eqs.~(\ref{8},\ref{14}) up to quartic order in
$\vec{\pi}$ and $\vec{\chi}$. In quadratic order,
discussed before, the minimizing $\pi$ configuration
in eq.~(\ref{8}) is linear in $\chi$:
\bee
\vec{\pi}_n=\sum_{n_B'} M(n,n_B')\vec{\chi}_{n_B'} \,,
\label{30}
\ee
where the matrix $M$ is short ranged (if $n$ is in the
block $n_B$ then only small $|n_B-n_B'|$ gives
significant contribution in the sum in eq.~(\ref{30}))
and can be obtained explicitly when solving the free field
problem \cite{14}. In order to obtain the $\chi$ dependence
of the right--hand side of eq.~(\ref{8}) up to quartic
order it is sufficient to substitute the minimizing
value of $\pi$ up to linear order, eq.~(\ref{30}).
The cubic corrections to eq.~(\ref{30}) give
O($\chi^6$) terms only in eq.~(\ref{8}).

In quartic order eq.~(\ref{8}) can be written in the form
\bee
\sum_{n_B,n_B',n_B'',n_B'''}
A(n_B,n_B';n_B'',n_B''')
\left( \vec{\chi}_{n_B}\vec{\chi}_{n_B'} \right)
\left( \vec{\chi}_{n_B''}\vec{\chi}_{n_B'''} \right)
=0 \,,
\label{31}
\ee
where the matrix $A$ depends linearly on $c(\ldots)$
and is symmetric under $n_B\leftrightarrow n_B'$,
$n_B''\leftrightarrow n_B'''$, and
$(n_B,n_B')\leftrightarrow (n_B'',n_B''')$.
Its form is not very illuminating and can be easily
derived, so we do not quote it here explicitly.
Although the matrix $A$ has no explicit $N$--dependence,
the solution $c(\ldots)$ distinguishes between $N\ge 3$
and $N=2$. For $N\ge 3$ eq.~(\ref{31}) implies $A=0$,
while for $N=2$~  $\chi$ is a 1--component vector
and eq.~(\ref{31}) gives
\bee
A(n_B,n_B';n_B'',n_B''')+
A(n_B'',n_B';n_B,n_B''')
=0 \,, ~~~~(N=2)\,.
\label{32}
\ee

These equations can be solved for $c(\ldots)$ by a
rapidly converging iterative procedure. The short range
nature of $\rho(r)$ for $\kappa=2$ is inherited by
$c(\ldots)$, as expected.
Table~2 enumerates those couplings
with $|c(\ldots)|>5\cdot 10^{-5}$.

\begin{table}  
\setlength{\unitlength}{0.8mm}
\begin{center}
\begin{tabular}{r cr c r cr}
\# & type & coupling & ~~~~~~ & \# & type & coupling  \\

1 &
\begin{picture}(20,10)(0,3)    %
\put(5,5){\circle*{2}}         
\put(15,5){\circle*{2}}        %
\put(5,4.5){\line(1,0){10}}
\put(5,5.5){\line(1,0){10}}
\end{picture} & 0.05344 & &

2 &
\begin{picture}(20,14)(0,5)    
\put(5,1){\circle*{2}}         
\put(15,11){\circle*{2}}       
\put(5,0.5){\line(1,1){10}}
\put(5,1.5){\line(1,1){10}}
\end{picture} & 0.00960  \\

3 &
\begin{picture}(20,14)(0,5)    
\put(5,1){\circle*{2}}         
\put(15,1){\circle*{2}}        
\put(15,11){\circle*{2}}
\put(5,1){\line(1,0){10}}
\put(15,1){\line(0,1){10}}
\end{picture} & 0.02155  & &

4 &
\begin{picture}(20,14)(0,5)    
\put(5,1){\circle*{2}}         
\put(15,1){\circle*{2}}        
\put(15,11){\circle*{2}}
\put(5,1){\line(1,0){10}}
\put(5,1){\line(1,1){10}}
\end{picture} & 0.01881   \\

5 &
\begin{picture}(20,14)(0,5)   
\put(5,1){\circle*{2}}        
\put(15,1){\circle*{2}}       
\put(15,11){\circle*{2}}      
\put(5,11){\circle*{2}}
\put(5,1){\line(1,1){10}}
\put(5,11){\line(1,-1){10}}
\end{picture} & 0.01209  & &

6 &
\begin{picture}(20,14)(0,5)   
\put(5,1){\circle*{2}}        
\put(15,1){\circle*{2}}       
\put(15,11){\circle*{2}}
\put(5,11){\circle*{2}}
\put(5,1){\line(1,0){10}}
\put(5,11){\line(1,0){10}}
\end{picture} & $-0.00258$  \\

7 &
\begin{picture}(20,14)(0,4)
\put(0,0){\circle*{2}}
\put(10,0){\circle*{2}}
\put(10,10){\circle*{2}}
\put(20,10){\circle*{2}}
\put(0,0){\line(2,1){20}}
\put(10,0){\line(0,1){10}}
\end{picture} & 0.00144  & &

8 &
\begin{picture}(20,14)(0,-1)   %
\put(0,0){\circle*{2}}         
\put(20,0){\circle*{2}}        %
\put(10,0){\circle*{2}}
\put(0,0){\line(1,0){20}}
\put(10,0){\line(1,0){10}}
\end{picture} & 0.00123  \\

9 &
\begin{picture}(20,14)(0,5)   
\put(0,0){\circle*{2}}        
\put(20,0){\circle*{2}}       
\put(10,1){\circle*{2}}       
\put(10,11){\circle*{2}}
\put(0,-0.5){\line(1,0){20}}
\put(10,1){\line(0,1){9}}
\end{picture} & 0.00121  & &

10 &
\begin{picture}(20,14)(0,4)
\put(0,0){\circle*{2}}
\put(10,0){\circle*{2}}
\put(10,10){\circle*{2}}
\put(20,10){\circle*{2}}
\put(0,0){\line(1,1){10}}
\put(10,0){\line(1,1){10}}
\end{picture} & $-0.00070$  \\

11 &
\begin{picture}(20,14)(0,4)
\put(0,0){\circle*{2}}
\put(10,10){\circle*{2}}
\put(20,0){\circle*{2}}
\put(0,0){\line(1,1){10}}
\put(10,10){\line(1,-1){10}}
\end{picture} & 0.00062  & &

12 &
\begin{picture}(20,14)(0,4)
\put(0,0){\circle*{2}}
\put(10,0){\circle*{2}}
\put(20,10){\circle*{2}}
\put(0,0){\line(1,0){10}}
\put(10,0){\line(1,1){10}}
\end{picture} & 0.00062  \\

13 &
\begin{picture}(20,14)(0,4)
\put(0,0){\circle*{2}}
\put(10,0){\circle*{2}}
\put(20,0){\circle*{2}}
\put(10,10){\circle*{2}}
\put(0,0){\line(1,0){10}}
\put(10,10){\line(1,-1){10}}
\end{picture} & $-0.00057$  & &

14 &
\begin{picture}(20,14)(0,4)
\put(0,0){\circle*{2}}
\put(0,10){\circle*{2}}
\put(10,0){\circle*{2}}
\put(20,10){\circle*{2}}
\put(0,0){\line(0,1){10}}
\put(10,0){\line(1,1){10}}
\end{picture} & $-0.00042$  \\

\end{tabular}
\end{center}
\caption{The coefficients of the leading quartic terms
$(1-\Sv_{n_1}\Sv_{n_2})(1-\Sv_{n_3}\Sv_{n_4})$ of
$\A^*$. A term which is defined by the sites
$n_1,n_2,n_3,n_4$ and the topology of the four--spin
coupling occurs only once with these coefficients in
the action. The numbers in this Table are multiples of
$c(n_1,\ldots,n_4)$ in eq.~(\protect\ref{14}).
This symmetry factor is 4 for the couplings \#1,2
and 8 for the others.}
\end{table}

\subsection{The numerical procedure to solve the FP
equation}

For a given configuration $\{\R\}$,
eqs.~(\ref{7},\ref{8}) can be solved numerically for
the FP value $\A^*(\R)$. Calculating $\A^*(\R)$
for a large number of different $\{\R\}$ configurations,
the important couplings (beyond those in $\rho(r)$ and
$c(\ldots)$) can be found by a fitting procedure.
The finite size distortion on the couplings will be small
if the lattice on which $\{\R\}$ is defined is larger
than the range of interaction in $\A^*$.
Since the later is small for $\kappa=2$ (see $\rho(r)$ and
$c(\ldots)$, and further discussion in 2.8) this
requirement is easy to satisfy. In this pilot study
we worked on $5\times 5$ periodic lattices when looking
for the couplings of the FP, $\A^*$ numerically.

We worked out two different numerical procedures to
calculate $\A^*(\R)$ for a given configuration
$\{\R\}$. Both were based on iterating eq.~(\ref{7}).
For a process with $k$ iterations we define the lattice
sizes $L^{(0)}$, $L^{(1)}=L^{(0)}/2$,\ldots,
$L^{(k)}=L^{(0)}/2^k$. The configuration $\{\R\}$
lives on the lattice of size $L^{(k)}$ (=5, in our case),
while we denote the spins on the $0^{th}$, $1^{st}$,
\ldots, $(k-1)^{th}$ level by
$\Sv^{(0)}_{n_0}$, $\Sv^{(1)}_{n_1}$,\ldots,
$\Sv^{(k-1)}_{n_{k-1}}$. We get
\beea
\A^{(k)}(\R)=\min_{ \{ \Sv^{(0)},\Sv^{(1)},\ldots,\Sv^{(k-1)}
\} }  & & \!\!\!\!\! \left\{ \A^{(0)}\left( \Sv^{(0)} \right)
-\kappa \sum_{n_1} \left[ \Sv^{(1)}_{n_1}
\sum_{n_0\in n_1} \Sv^{(0)}_{n_0} -
| \sum_{n_0\in n_1} \Sv^{(0)}_{n_0} | \right] \right.
\nonumber \\
& & ~~~~~~~~~~~~~~\, -\kappa \sum_{n_2} \left[ \Sv^{(2)}_{n_2}
\sum_{n_1\in n_2} \Sv^{(1)}_{n_1} -
| \sum_{n_1\in n_2} \Sv^{(1)}_{n_1} | \right]
\label{34} \\
& & \vdots \nonumber \\
& & ~~ \left. -\kappa \sum_{n_k} \left[ \R_{n_k}
\sum_{n_{k-1}\in n_k} \Sv^{(k-1)}_{n_{k-1}} -
| \sum_{n_{k-1}\in n_k} \Sv^{(k-1)}_{n_{k-1}} | \right]
\right\} \,. \nonumber
\eea
Here $\A^{(0)}$ is the zeroth order approximation
for $\A^*$, while $\A^{(k)}(\R)$ is the $k^{th}$
approximation for the number $\A^*(\R)$.
For $\A^{(0)}$ one might take even the standard action,
but the convergence in $k$ will improve if a better
approximation is used. (For example, one can use
the results on $\rho(r)$ and $c(\ldots)$, or those
of earlier numerical runs.)
In eq.~(\ref{34}) $\sum_{n_0\in n_1} \Sv^{(0)}_{n_0}$
denotes a sum over the 4 spins in the block $n_1$, etc.

Note that eq.~(\ref{34}) and the way we solved it
are strongly reminiscent to a multigrid procedure \cite{Ma,17}.

In the first method the minimalization in $\Sv$ was
performed locally and the lattices were swept through
until $\A^{(k)}(\R)$ became stable to a given precision.
In the second method we used annealed cooling.
The largest irrelevant eigenvalue of the FP is $1/4$
in agreement with the observed convergence rate in the
number of blocking steps $k$.
Even taking the crude approximation for $\A^{(0)}$
where only the nearest neighbour and diagonal
couplings from $\rho(r)$ are kept,
$|\A^{(4)}(\R)-\A^*(\R)| / |\A^*(\R)|$ was estimated
less than O($10^{-4}$).
The two different methods gave consistent results.

\subsection{The properties of $\A^*$ on coarse
configurations}

On smooth configurations the first terms in eq.~(\ref{14})
dominate. These terms are short ranged and explicitly
known. It is important to investigate, however, whether
$\A^*$ remains short ranged even on coarse configurations.
We need further hints on the general parametrization
problem also.

In order to see how $\A^*(\Sv)$ behaves as the relative angle
between the spins is increased, we rotated a single spin
$\Sv$ away from the $1^{st}$ axis by an angle $\vartheta$,
while all the other spins pointed in the $1^{st}$ direction.
The corresponding value of  $\A^*$,
which we denote by $\A_1(\vartheta)$, is shown in Fig.~6.
The curve is very simple and can be fitted easily if we
parametrize it in powers of $\vartheta^2/2$ rather than in powers
of $(1-\cos\vartheta)$. This empirical observation is also
supported by considering a quasi one--dimensional solution
to the equations of motion, with
$\Sv(n)=(\cos\vartheta n_0,\sin\vartheta n_0,0)$,
where the $\vartheta$ dependence of $A^*$ is
{\it exactly} given by
$\vartheta^2/2$. So, we rewrite eq.~(\ref{14}) as
$$
\A(\Sv)=-\frac{1}{2}\sum_{n,r}\rho(r) \frac{1}{2}
\vartheta_{n,n+r}^2
+\sum_{n_1,n_2,n_3,n_4} \bar{c}(n_1,n_2,n_3,n_4)
\frac{1}{2}\vartheta_{n_1,n_2}^2 \cdot
\frac{1}{2}\vartheta_{n_3,n_4}^2 +\ldots
\eqno{(12')}
$$
where $\vartheta_{n_1,n_2}$ is the angle between the spins
$\Sv_{n_1}$ and  $\Sv_{n_2}$. The function $\bar{c}$ is trivially
related to $\rho$ and $c$ of eq.~(\ref{14}).
Using this parametrization, the leading couplings of
$\rho$ and $\bar{c}$ will dominate $\A^*$
even on coarse configurations. This is illustrated in Fig.~6,
where a $d_1 \vartheta^2/2 + d_2 (\vartheta^2/2)^2$ form, with
$d_1$ and $d_2$ obtained from $\rho$ and $c$ in Tables~1,2,
agree very well with the data.

Next we wanted to check whether the distant couplings remain
small even at large relative angles. In a trivial background
(spins in the $1^{st}$ direction) we rotated $2$ spins, which
were at a distance $\Delta$ from each other, by an angle $\vartheta$
relative to the $1^{st}$ axes (keeping them parallel to each
other). Denoting the corresponding value of $A^*$ by
$A_2(\vartheta)$  it is easy to see that
$A_1(\vartheta)-\frac{1}{2} A_2(\vartheta)$ is a measure of the
direct coupling between the two spins. Table~3 gives the measured
values at different $\vartheta$ values for distance
$\Delta=(2,0)$ compared
with $-\rho(2,0)\, \vartheta^2/2$, the first, exactly known
term in eq.~(12').
The direct coupling remains small and very well given
by this single term even at large angles.

\begin{table} 
\begin{center}
\begin{tabular}{|c|c|c|c|c|}
\hline
$\vartheta$ & $\A_1(\vartheta)$ & $\A_2(\vartheta)$ &
$\A_1(\vartheta)-\frac{1}{2}\A_2(\vartheta)$ &
$-\rho(2,0)\vartheta^2/2$ \\
\hline
0.09967 & 0.016094 & 0.032166 & 0.000011 & 0.000010 \\
$\pi/4$ & 0.99558 & 1.98986 & 0.00065 & 0.00062 \\
$\pi/2$ & 3.9339  & 7.8627 & 0.0025 & 0.0025 \\
$2.678$ & 10.9999 & 21.9863 & 0.0067 & 0.0072 \\
$2.85$ & 12.3352 & 24.6553 & 0.0075 & 0.0081 \\
\hline
\end{tabular}
\end{center}
\caption{One and two spins (at a separation (2,0)) are rotated
in a trivial background. The values of $\A^*$ for the corresponding
configurations are given together with
$\A_1(\vartheta)-\frac{1}{2}\A_2(\vartheta)$ which measures
the direct (2,0) interaction. The last column shows the
contribution from the first term in eq.~(12').}
\end{table}

\subsection{Fixing the couplings in this pilot study
in O(3)}
In the following we consider the O(3) model, $N=3$. We
generated $\sim 500$ different ${\R}$ configurations
and calculated the corresponding values of the FP, $\A^*(\R)$
using the minimalization procedure discussed in 2.7.
Among these configurations there were smooth,
but also quite coarse ones, typical for a correlation length
of a few lattice units only. We made fits with up to $69$
different couplings, taking them all short ranged,
with the largest distance being $(2,2)$.

Fig.~7 shows a fit with $24$ couplings, $8$ fixed analytically by
$\rho$ and $c$, 16 fitted. The couplings and their fitted values
are given in Table~4. The couplings \#1 and 4 are slightly
renormalized to satisfy the condition
$\sum_r \rho(r) r^2=-4$ exactly. In this fit all the couplings
can be put on a $1\times 1$ square. In Table~4 the couplings are
given in a graphical notation. The coupling \#15, for example,
multiplies the expression
${1\over 2}\vartheta^2_A\cdot {1\over 2}\vartheta^2_B \cdot
({1\over 2}\vartheta^2_C)^2$, where
$\vartheta_A=\acos(\Sv_1 \Sv_2)$,
$\vartheta_B=\acos(\Sv_1 \Sv_3)$
and $\vartheta_C=\acos(\Sv_2 \Sv_3)$. Here $\Sv_1,\Sv_2,\Sv_3$
are the $3$ spins forming the triangle, $\Sv_1$ sitting at the
$90^\circ$ corner.

\begin{table}  
\setlength{\unitlength}{0.8mm}
\begin{tabular}{r cr c cr c cr}
\# & type & coupling & ~~~ & type & coupling
& ~~~ & type & coupling \\

1 &
\begin{picture}(20,10)(0,3)
\put(5,5){\circle*{2}}
\put(15,5){\circle*{2}}
\put(5,5){\line(1,0){10}}
\end{picture} & $0.61884$ & &
\begin{picture}(20,10)(0,3)
\put(5,5){\circle*{2}}
\put(15,5){\circle*{2}}
\put(5,4.5){\line(1,0){10}}
\put(5,5.5){\line(1,0){10}}
\end{picture} & $-0.04957$ & &
\begin{picture}(20,10)(0,3)
\put(5,5){\circle*{2}}
\put(15,5){\circle*{2}}
\put(5,4.1){\line(1,0){10}}
\put(5,5){\line(1,0){10}}
\put(5,5.9){\line(1,0){10}}
\end{picture} & $-0.01163$ \\

4 &
\begin{picture}(20,14)(0,5)
\put(5,1){\circle*{2}}
\put(15,11){\circle*{2}}
\put(5,1){\line(1,1){10}}
\end{picture} & $0.19058$  & &

\begin{picture}(20,14)(0,5)
\put(5,1){\circle*{2}}
\put(15,11){\circle*{2}}
\put(5,0.5){\line(1,1){10}}
\put(5,1.5){\line(1,1){10}}
\end{picture} & $-0.02212$  & &

\begin{picture}(20,14)(0,5)
\put(5,1){\circle*{2}}
\put(15,11){\circle*{2}}
\put(5,0.1){\line(1,1){10}}
\put(5,1.9){\line(1,1){10}}
\put(5,1){\line(1,1){10}}
\end{picture} & $-0.00463$ \\

7 &
\begin{picture}(20,14)(0,5)
\put(5,1){\circle*{2}}
\put(15,1){\circle*{2}}
\put(15,11){\circle*{2}}
\put(5,1){\line(1,0){10}}
\put(5,1){\line(1,1){10}}
\end{picture} & $0.01881$  & &

\begin{picture}(20,14)(0,5)
\put(5,1){\circle*{2}}
\put(15,1){\circle*{2}}
\put(15,11){\circle*{2}}
\put(5,0.5){\line(1,0){10}}
\put(5,1.5){\line(1,0){10}}
\put(5,1){\line(1,1){10}}
\end{picture} & $-0.00139$  & &

\begin{picture}(20,14)(0,5)
\put(5,1){\circle*{2}}
\put(15,1){\circle*{2}}
\put(15,11){\circle*{2}}
\put(5,1){\line(1,0){10}}
\put(5,0.5){\line(1,1){10}}
\put(5,1.5){\line(1,1){10}}
\end{picture} & $0.00497$ \\

10 &
\begin{picture}(20,14)(0,5)
\put(5,1){\circle*{2}}
\put(15,1){\circle*{2}}
\put(15,11){\circle*{2}}
\put(5,1){\line(1,0){10}}
\put(15,1){\line(0,1){10}}
\end{picture} & $0.02155$  & &

\begin{picture}(20,14)(0,5)
\put(5,1){\circle*{2}}
\put(15,1){\circle*{2}}
\put(15,11){\circle*{2}}
\put(5,0.5){\line(1,0){10}}
\put(5,1.5){\line(1,0){10}}
\put(15,1){\line(0,1){10}}
\end{picture} & $0.00717$  & &

\begin{picture}(20,14)(0,5)
\put(5,1){\circle*{2}}
\put(15,1){\circle*{2}}
\put(15,11){\circle*{2}}
\put(5,0.5){\line(1,0){10}}
\put(5,1.5){\line(1,0){10}}
\put(14.5,1){\line(0,1){10}}
\put(15.5,1){\line(0,1){10}}
\end{picture} & $-0.00055$ \\

13 &
\begin{picture}(20,14)(0,5)
\put(5,1){\circle*{2}}
\put(15,1){\circle*{2}}
\put(15,11){\circle*{2}}
\put(5,1){\line(1,0){10}}
\put(5,1){\line(1,1){10}}
\put(15,1){\line(0,1){10}}
\end{picture} & $0.01078$  & &

\begin{picture}(20,14)(0,5)
\put(5,1){\circle*{2}}
\put(15,1){\circle*{2}}
\put(15,11){\circle*{2}}
\put(5,0.5){\line(1,0){10}}
\put(5,1.5){\line(1,0){10}}
\put(5,1){\line(1,1){10}}
\put(15,1){\line(0,1){10}}
\end{picture} & $0.00765$  & &

\begin{picture}(20,14)(0,5)
\put(5,1){\circle*{2}}
\put(15,1){\circle*{2}}
\put(15,11){\circle*{2}}
\put(5,1){\line(1,0){10}}
\put(5,0.5){\line(1,1){10}}
\put(5,1.5){\line(1,1){10}}
\put(15,1){\line(0,1){10}}
\end{picture} & $-0.00557$ \\

16 &
\begin{picture}(20,14)(0,5)
\put(5,1){\circle*{2}}
\put(15,1){\circle*{2}}
\put(15,11){\circle*{2}}
\put(5,11){\circle*{2}}
\put(5,1){\line(1,1){10}}
\put(5,11){\line(1,-1){10}}
\end{picture} & $0.01209$  & &

\begin{picture}(20,14)(0,5)
\put(5,1){\circle*{2}}
\put(15,1){\circle*{2}}
\put(15,11){\circle*{2}}
\put(5,11){\circle*{2}}
\put(5,0.5){\line(1,1){10}}
\put(5,1.5){\line(1,1){10}}
\put(5,11){\line(1,-1){10}}
\end{picture} & $-0.00114$  & &

\begin{picture}(20,14)(0,5)
\put(5,1){\circle*{2}}
\put(15,1){\circle*{2}}
\put(15,11){\circle*{2}}
\put(5,11){\circle*{2}}
\put(5,0.5){\line(1,1){10}}
\put(5,1.5){\line(1,1){10}}
\put(5,10.5){\line(1,-1){10}}
\put(5,11.5){\line(1,-1){10}}
\end{picture} & $0.00548$ \\

19 &
\begin{picture}(20,14)(0,5)
\put(5,1){\circle*{2}}
\put(15,1){\circle*{2}}
\put(15,11){\circle*{2}}
\put(5,11){\circle*{2}}
\put(5,1){\line(1,0){10}}
\put(5,11){\line(1,0){10}}
\end{picture} & $-0.00258$  & &

\begin{picture}(20,14)(0,5)
\put(5,1){\circle*{2}}
\put(15,1){\circle*{2}}
\put(15,11){\circle*{2}}
\put(5,11){\circle*{2}}
\put(5,0.5){\line(1,0){10}}
\put(5,1.5){\line(1,0){10}}
\put(5,11){\line(1,0){10}}
\end{picture} & $0.00387$  & &

\begin{picture}(20,14)(0,5)
\put(5,1){\circle*{2}}
\put(15,1){\circle*{2}}
\put(15,11){\circle*{2}}
\put(5,11){\circle*{2}}
\put(5,0.5){\line(1,0){10}}
\put(5,1.5){\line(1,0){10}}
\put(5,10.5){\line(1,0){10}}
\put(5,11.5){\line(1,0){10}}
\end{picture} & $-0.00100$ \\

22 &
\begin{picture}(20,14)(0,5)
\put(5,1){\circle*{2}}
\put(15,1){\circle*{2}}
\put(15,11){\circle*{2}}
\put(5,11){\circle*{2}}
\put(5,1){\line(1,0){10}}
\put(5,11){\line(1,0){10}}
\put(5,1){\line(0,1){10}}
\end{picture} & $-0.01817$  & &

\begin{picture}(20,14)(0,5)
\put(5,1){\circle*{2}}
\put(15,1){\circle*{2}}
\put(15,11){\circle*{2}}
\put(5,11){\circle*{2}}
\put(5,1){\line(1,0){10}}
\put(5,11){\line(1,0){10}}
\put(4.5,1){\line(0,1){10}}
\put(5.5,1){\line(0,1){10}}
\end{picture} & $-0.00772$  & &

\begin{picture}(20,14)(0,5)
\put(5,1){\circle*{2}}
\put(15,1){\circle*{2}}
\put(15,11){\circle*{2}}
\put(5,11){\circle*{2}}
\put(5,1){\line(1,0){10}}
\put(5,11){\line(1,0){10}}
\put(5,1){\line(0,1){10}}
\put(15,1){\line(0,1){10}}
\end{picture} & $0.04970$  \\

\end{tabular}
\caption{The couplings used for the FP--action in eq.~(12').
The notation is explained in the text. The coefficients
of the quadratic and quartic interactions are calculated
analytically, the higher order interactions are fitted.}
\end{table}

As Fig.~7 shows the fit becomes poorer on coarse configurations
(those have larger action values), in a few cases the error
reaches $\sim 1\%$, but for most of the configurations the error
is much smaller. Including the $(2,0),(2,1)$ and $(2,2)$
couplings the fit becomes somewhat better, but for this pilot
study we decided to use the $24$--parameter fit in Table~4.

We want to emphasize two points here. Using $24$, or even twice
as many couplings in a Monte Carlo simulation is not really a
problem -- at least in the spin model considered here.
Although this action includes 3-- and 4--spin interactions
also, one can generalize the cluster Monte Carlo technique
\cite{18} for this case.
Our cluster program runs $\sim 3$ times slower with $24$
parameters than with the standard action. Relative to the gain
we are looking for, this is irrelevant. The second remark concerns
the numbers in Table~4. The fitted parameters are effective values,
one must not associate significance to a single entry in this Table.
Modifying somewhat the set of operators used in the fit,
these entries would change, but the global quality of the
fit would hardly be different.

\subsection{Results obtained by simulating the FP action}

The FP--action $\A^*_{FP}$ is defined as $\beta_{FP}\A^*(\Sv)$.
The coupling $\beta_{FP}$
is fixed uniquely by the normalization condition on $\rho$
(sect.~2.4). For large $\beta_{FP}$, the RT runs together with the
FP--action (Fig.~2), therefore  $\A^*_{FP}$ is the perfect action.
For intermediate  $\beta_{FP}$ values it is not perfect anymore,
but -- as we shall see -- it performs amazingly well. As we
discussed in the previous sections, the FP--action can be determined
with the help of classical calculations. For $\kappa=2$ it is
very short ranged and a relatively simple parametrization describes
it well. This parametrized form (given in Table~4) can
be easily simulated. Our cluster Monte Carlo program performed very
effectively. We would like to urge workers in this field to replace
the standard action by the FP--action as a first approximation
towards 'perfectness'. These remarks apply to the $CP^n$--model
also, especially in studies related to topology (sect.~2.2).

In the following, for simplicity, we shall call the $24$--parameter
action in Table~4 the FP--action keeping in mind possible effects
from the errors of this simple parametrization. In the Appendix
we calculated the relation of the $\Lambda$--parameter of the
FP--action to that of the standard action. The result
\bee
{\Lambda_{FP}\over \Lambda_{ST}}=8.17\,, ~~~~~~
\mbox{(N=3, couplings from Table~4)}\,.
\label{35}
\ee
can also be written as
\bee
\beta_{FP}=\beta_{ST}-0.334+ {\rm O}(1/\beta_{ST})\,.
\label{36}
\ee
Eq.~(\ref{36}) gives a feeling about the meaning of $\beta_{FP}$.
At intermediate $\beta$ values we give --- for orientation ---
the infinite volume correlation length $\xi(\beta_{FP})$ at a
few $\beta_{FP}$ values: $\xi(1.214)\approx 36$,
$\xi(1.08)\approx 18$, $\xi(0.85)\approx 6$.

\subsubsection{Cut--off dependence of the running coupling}
We discussed this problem in the Introduction already.
We have chosen the $g(L)=m(L)L=1.0595$ point from ref.~\cite{2}
since here the observed cut--off effects in the relation $m(2L)2L$
versus $m(L)L$ were large and opposite in sign with respect to
perturbation theory (Fig.~1). We simulated the FP--action on a
lattice of spatial size $L/a=5$ and tuned $\beta_{FP}$ until
$m(L)L$ became close to the prescribed value:
at $\beta_{FP}=1.0821$ we got $g(L)=1.0578(5)$.
Then measuring on a lattice with $L'/a=2L/a=10$
at the same $\beta_{FP}$ value we had $m(2L)2L = 1.2611(9)$.
In the time direction the lattice
was chosen to be at least $6$--times larger than the finite box
correlation length $\xi(L)=1/m(L)$ and distances larger than
three times $\xi$ were used in the fitting procedure to obtain
the mass gap. At the end we shifted $g(2L)$ according to the
slight difference between the actual (1.0578) and
prescribed (1.0595) value for $g(L)$ obtaining $g(2L)=1.2638(12)$.
For this shift we used the universal curve obtained in \cite{2}.
The error from this procedure is negligible.
We repeated this calculation at $L=10$ also, for $\beta_{FP}=1.214$
with the results: $m(L)L=1.0613(8)$ and $m(2L)2L=1.2664(18)$.
After shifting  this gives $g(2L)=1.2635(22)$.
In Fig.~3 these two points are compared with the extrapolated
prediction $g(2L)=1.2641(20)$.
No cut--off effects can be seen.\footnote{Surprisingly,
even for an extremely coarse lattice, $L/a=3$,
(at $\beta_{FP}=0.98$) one obtains $g(2L)=1.2626(11)$,
still with no sign of lattice artefacts.}

\subsubsection{The two--point function}
We want to compare the spin--spin correlation function of the
standard nearest neighbour model with that of the FP--action
on coarse lattices and check the extent of violation of rotation
symmetry. In this respect one has to understand the following
point. The correlation functions do not describe directly physics
since they depend on the way the fields are defined. This is
reflected, in general, in the scheme dependence of the
wave--function renormalization. Here the relation between the
fields of the FP--action and the perfect fields of the
continuum is more involved. In the case of a free field, as the
derivation in section 2.4 (especially eq.~(\ref{23})) clearly
shows, the field of the FP--action is the integral of the perfect
field over a square of length $a$. This definition brings an
inherent rotation symmetry breaking in the two--point function.
Similarly, in the full O(3) model the shape of the block
used in the RG transformation gives a non--rotation invariant
definition for the fields in the FP--action. Of course, physical
predictions, like the energy values related to the
exponential decay of the two--point function, are rotation
symmetric.

The effect discussed above is, however, very small. This can be
checked analytically in the free field case and it remains so
in the O(3) non--linear $\sigma$--model even on coarse
lattices. We simulated the FP--action and the standard action
at correlation length $\sim 3$
($\beta_{FP}$=0.7 and $\beta_{ST}$=1.18, respectively). We have
calculated the two--point function on a $24\times 24$ periodic
lattice. This lattice is large enough to avoid rotation symmetry
breaking from the box itself (which is a physical infrared
effect) up to distances $\sim 9$. In Figs.~8,9 the measured
points are connected by piece--wise straight lines to lead
the eyes. No symmetry breaking effects can be seen in the
two--point function of the FP--action, while the lattice structure
shows up clearly when the standard action is simulated.

\section{RG transformation at finite $\beta$}

The FP--action is the perfect classical action, but we need
the perfect quantum action. Since the FP--action seems to have
small cut--off effects even at intermediate $\beta$ values,
it is a good starting point to find the RT.
Assume that at some $\beta_{FP}$ value the cut--off effects
in the predictions of the FP--action $\beta_{FP}\A^*(\Sv)$
are negligible. Performing a RG step, the size of the cut--off
effects for the blocked action will be the same (i.e. negligible),
but the correlation length will be a factor of 2 smaller.
This is the basic step we have to do in searching for the
perfect action.
In the following we shall illustrate that it is relatively
easy to determine and parametrize the blocked action,
and this blocked action remains short ranged if the RG
transformation is chosen properly.

\subsection{The RG transformation}

We shall perform one RG step starting with a FP--action
(in its parametrical form as given in Table~4) at
$\beta_{FP}=1.0$. At this point the correlation length
is O(10). We use the block transformation defined in
eq.~(\ref{3}). There is no reason to assume that the optimal
value of the parameter $P$ is $\beta_{FP}\cdot 2$ as it is
in the case for large $\beta_{FP}$. Actually, we expect that
$P_{opt}$ goes to a constant rather than to zero for small
\(\beta_{FP}\).
Preliminary runs indicated that $\kappa=2.5$ is close to the
optimal choice, so we took $P=\beta_{FP}\cdot 2.5$.
In this case the range of interaction of the blocked action
turned out to be essentially the same as that of the
FP--action. For this reason it is sufficient to start with
a $10\times 10$ lattice and block it down to $5\times 5$.
At this small lattice sizes the Ferrenberg--Swendsen \cite{19}
technique becomes especially effective and we used it
extensively.

\subsection{Determination of the properties of the blocked
action}

In a MC calculation it is relatively easy to determine
the {\it change} of the action under the change of the
configuration. This information will be enough for us.
Let us introduce the notation
\bee
T(\R,\Sv)=\sum_{n_B}\left[ P \, \R_{n_B}
\sum_{n\in n_B}\Sv_n -\ln Y_N\left(
P | \sum_{n\in n_B}\Sv_n | \right) \right] \,.
\label{37}
\ee
Eq.~(\ref{3}) gives then
\bee
\rme^{-\beta' \left[ \A'(\R)-\A'(\R_0) \right] }
=\left\langle \rme^{T(\R,\Sv)-T(\R_0,\Sv)}
\right\rangle_{\R_0} \,,
\label{38}
\ee
where
\bee
\left\langle {\cal O} \right\rangle_{\R_0}
= {\int_{\Sv} \rme^{-\beta_{FP}\A^*(\Sv)+T(\R_0,\Sv)}
{\cal O} \over
\int_{\Sv} \rme^{-\beta_{FP}\A^*(\Sv)+T(\R_0,\Sv)} } \,,
\label{39}
\ee
and $\{\R\}$ and $\{\R_0\}$ are two different
block--configurations.

The difference  $T(\R,\Sv)-T(\R_0,\Sv)$ depends on the
block--averages
\bee
{\bf f}_{n_B}=\sum_{n\in n_B} \Sv_n \,.
\label{40}
\ee

Simulating the effective action defined in eq.~(\ref{39})
with some fixed $\{\R_0\}$ and storing $\{{\bf f}_{n_B}\}$
after every sweep, the expectation value in eq.~(\ref{38})
can be calculated for any configuration $\{\R\}$.
In practice $\{\R\}$ should be close to $\{\R_0\}$,
otherwise the statistical errors can not be controlled.

Specifically, one can chose $\{\R_0\}$ to be the trivial
configuration (all the spins point in the $1^{st}$ direction).
In this case $\A'(\R_0)=0$ (by definition; the constant
generated by the RG transformation does not interest us),
and we can calculate $\A'(\R)$ for configurations
which are close to the trivial one.
By taking a parametrization for $\A'$ in the form of
eq.~(12') we get
\bee
\beta'\rho'(r)= P^2 Q^{\bot}(r),~~~~ r\neq 0 \,,
\label{41}
\ee
where $(N=3)$
\bee
Q^{\bot}(n_B'-n_B)={1\over 2}
\left\langle f_{n_B}^2 f_{n_B'}^2+f_{n_B}^3 f_{n_B'}^3
\right\rangle_{\R_0=triv} \,.
\label{42}
\ee
The conventions
\bee
\sum_r \rho'(r)=0,~~~~ \sum_r \rho'(r) r^2= -4
\label{43}
\ee
determine then $\rho'(0)$ and $\beta'$, respectively.
We summarized the results in Table~5. It is interesting to
observe that the structure of $\rho'(r)$ is quite similar to
$\rho(r)$. Most importantly, it is similarly short ranged.

\begin{table}  
\begin{center}
\begin{tabular}{|c|c||c|c|}
\hline
$r$ & $\beta'\rho'(r)$ & $r$ & $\beta'\rho'(r)$ \\
\hline
(1,0) & 0.549(1) & (2,0) & 0.0037(14) \\
(1,1) & 0.167(1) & (2,1) & 0.0027(13) \\
      &          & (2,2) & 0.0002(13) \\
\hline
\end{tabular}
\end{center}
\caption{After one RG step on the FP--action at $\beta=1.0$
the new action $\beta'\A'$ is parametrized according to
eq.~(12'). The leading couplings in $\beta'\rho'(r)$
are given above. The conventions in eq.~(\protect\ref{43})
give $\beta'=0.92(2)$.}
\end{table}

We can now take configurations $\{\R\}$, where a single spin
is rotated by an angle $\vartheta$, in a trivial background.
The corresponding value of $\beta'\A'$ is given in the second
column of Table~6. As in the case of the FP, $\A^*$, one observes
a simple power--like behaviour in $\vartheta$, which justifies
a parametrization in the form of eq.~(12'). Actually, the
two important couplings in $\rho'$ which give a contribution
$1.432 \,\vartheta^2$ (third column in Table~6) dominate
the results even at large angles $\vartheta$. When two
spins, separated by the vector $(2,0)$, are rotated together,
the measured action (column 4 in Table~6) is very closely
twice of the result for a single rotated spin.
The direct $(2,0)$ coupling is small.

\begin{table}   
\begin{center}
\begin{tabular}{|c|c|c||c|c|}
\hline
  & \multicolumn{2}{c|}{one spin rotated} &
  \multicolumn{2}{c|}{two spins rotated} \\
\cline{2-5}
$\vartheta$ & $\beta'\A'$ & $1.432\,\vartheta^2$ & $\beta'\A'$ &
direct (2,0) interaction \\
\hline
0.09967 & 0.01427(1) & 0.01423 & 0.02850(2) & 0.000037(28) \\
$\pi/4$ & 0.8780(7) & 0.8833 & 1.753(2) & 0.0029(23) \\
$3\pi/8$ & 1.953(3) & 1.987 & 3.91(1)  & 0.000(14) \\
$\pi/2$ & 3.42(1)  & 3.53 & 6.91(6)  & $-0.08(6)$ \\
$5\pi/8$ & 5.22(3) & 5.52 &   &  \\
$3\pi/4$ & 7.26(6) & 7.95 &   &  \\
2.8  & 9.31(9) & 11.23 &   &  \\
3.0  & 9.8(1) & 12.89 &   &  \\
\hline
\end{tabular}
\end{center}
\caption{One and two spins (at a separation (2,0)) are rotated
in a trivial background. The two--spin result ($4^{th}$ column)
is almost exactly twice of the one spin result, indicating that
the direct (2,0) interaction is small. The two leading couplings
of $\rho'$, giving $1.432 \,\vartheta^2$, dominate the action
even at large angles.}
\end{table}

Since $\A'$ is short ranged, for the general parametrization
we might consider the type of couplings used for the
FP--action before. Actually, as we shall see, the  FP--action
(as parametrized in Table~4.) $\beta_{FP}'\A^*$ describes
$\beta'\A'$ quite well with $\beta_{FP}'\sim 0.85$.
In applications the blocked action $\beta'\A'$ will be
used in a MC simulation, therefore we want a parametrization
which gives the value of the action precisely for typical
configurations. We generated 300 configurations by MC using
the FP--action with $\beta_{FP}'=0.85$. We calculated the
value of the blocked action $\beta'\A'$ on these configurations
using the technique discussed before.
The type of the couplings kept for the fit where those
used for the FP--action, plus the $(2,0)$, $(2,1)$ and $(2,2)$
two--spin couplings and their powers up to $\vartheta^6$.
In order to see the quality of the fit and to separate
in the fitting error the statistical (coming from MC
determination of the action) and systematical (coming
from parametrization) errors we kept those configurations
only where the statistical error was below a certain cut value.
Fig.~10 shows the error of the fit against the value
of the action. The average value of the fit error is $0.004$.
If the $(2,0)$, $(2,1)$ and $(2,2)$ couplings are left out,
the average fit error is increased to $0.010$.
If needed, the quality of the fit can be increased by including
further couplings (and increasing the statistical precision).
In our feasibility study we did not investigate this
problem further.
We want to mention two additional points only.
The FP--action $\beta_{FP}'\A^*$ with $\beta_{FP}'=0.848$
gives a quite reasonable fit also with an average fit--error
of 0.022. The second remark concerns the value $\kappa=2.5$
which makes the blocked action somewhat more local and
closer to the FP--action than $\kappa=2$ we used at
$\beta=\infty$. At $\kappa=2$, for example,
$\beta'\rho'(2,0)=0.011(1)$, a factor of $\sim 3$ larger
than at $\kappa=2.5$. Similarly, the fit--errors (analogous
to those discussed above) are increased by a factor of
$\sim 2$ if $\kappa=2$ is taken.

\noindent{\it Acknowledgement.} The authors are indebted for the
useful discussions to  M. Blatter, R. Burkhalter,  G. Mack,
K.-H. M\"utter, A. Pordt, K. Schilling, P. Weisz, U. Wolff
and U. Wiese. We thank M. L\"uscher for raising our attention
to a minor flaw in the proof of the statement on classical solutions
in the preprint version of this paper.

\appendix
\section{The Lambda parameter}

The $\Lambda$ parameter for some specific lattice
regularizations (in units of $\Lambda_{ST}$ corresponding
to the standard action, for example) has been obtained
earlier \cite{20,21}. We give here the result for the generic case,
for an action of the form given by eq.~(\ref{14}).

The simplest way to calculate the ratio of the
$\Lambda$ parameters is to introduce a chemical potential
$h$ \cite{22} coupled to a Noether charge of the O($N$) symmetry
and use the fact that the free energy $f(h)$ depends
on $h$ in a universal way \cite{22}.

Technically, the chemical potential is a constant, imaginary
gauge potential: $A_{\mu} \to i h \delta_{\mu 0} q$,
where $q$ is a generator of an O($N$) rotation. We choose
a rotation in the 1,2 plane:
\bee
q=\left( \begin{array}{cccc}
                             0 & -i & & 0 \\
                             i & 0  & & 0 \\
                               &   & \ddots & 0 \\
                             0 & 0  & & 0
         \end{array} \right)
\label{q}
\ee

In the presence of the chemical potential the only change
in the action is in the scalar product of two spins:
\beea
& & \Sv_n \Sv_{n+r} \to \Sv_n \rme^{hr_0 q} \Sv_{n+r}
\approx \Sv_n \Sv_{n+r}  \label{sprodh} \\
& & ~~ + ihr_0 \left(\Sv_n^{(1)} \Sv_{n+r}^{(2)}-
\Sv_n^{(2)} \Sv_{n+r}^{(1)} \right)
+{1\over 2} h^2 r_0^2 \left(\Sv_n^{(1)} \Sv_{n+r}^{(1)}+
\Sv_n^{(2)} \Sv_{n+r}^{(2)} \right)+\ldots \nonumber
\eea
(In fact, $h$ is here $ah$ and  the omitted terms
 do not contribute in the continuum limit.)

Using the representation in  eq.~(\ref{16}) and expanding
in $\vec{\pi}$, after straightforward calculations
we obtain the following expression for the
free energy:
\beea
 & & \!\!\!\!f(h)= -{h^2\over 2g}+
{1\over 2}\int_q \ln \left\{
\tilde{\rho}(q) +h^2 +{1\over 2} h^2 \sum_r \rho(r) r_0^2
\cos qr - h^2\!\!\!\sum_{n_1,\ldots,n_4}\!\!\!
c(n_1,n_2;n_3,n_4) \right. \nonumber \\
 & &
\!\!\times \left. \left[
8 r_0 r_0' \cos qd \sin{qr\over 2}\sin{q'r\over 2}
+ r_0^2 (1-\cos qr')+ {r_0'}^2 (1-\cos qr)
\right] \right\}  \label{fh} \\
& &\!\!\!\! +{1\over 2}(N-2) \int_q \ln \left\{
\tilde{\rho}(q) +h^2
-h^2\!\!\!\sum_{n_1,\ldots,n_4}\!\!\! c(\ldots)
\left[
r_0^2 (1-\cos qr')+ {r_0'}^2 (1-\cos qr)
\right] \right\}  \,.
\nonumber
\eea

Expanding in $h$ and using the fact that $f(h)-f(0)$ is
the same for any two regularizations, one obtains:
\bee
-{1\over g}+Q=-{1\over g'}+Q'
\label{inv}
\ee
where
\beea
Q= & & \int_q{1\over \tilde{\rho}(q)} \left\{ N-2
-{1\over 4}\sum_r \rho(r) r^2 (1-\cos qr)\right.\nonumber \\
& & -4\!\!\!\sum_{n_1,\ldots,n_4} \!\!\!c(n_1,n_2;n_3,n_4)
(r\cdot r')\cos qd \sin{qr\over 2}\sin{q'r\over 2}
 \label{Q} \\
& & \left. -{1\over 2}(N-1)\!\!\!\sum_{n_1,\ldots,n_4}
\!\!\! c(n_1,n_2;n_3,n_4)
\left[ r^2(1-\cos qr')+ {r'}^2 (1-\cos qr) \right]
\right\} \,.
\nonumber
\eea
Here we introduced the notations for the relative
coordinates in $c(n_1,n_2;n_3,n_4)$:
\bee
r=n_1-n_2,~~~r'=n_3-n_4,~~~ d={1\over 2}(n_1+n_2-n_3-n_4) \,,
\label{relc}
\ee
and used the $90^\circ$ symmetry to replace e.g. $2 r_0 r_0'$
by the scalar product $(r\cdot r')$.

Eq.~(\ref{inv}) gives for the ratio of two $\Lambda$
parameters:
\bee
{\Lambda'\over \Lambda}=\exp\left\{-{2\pi\over N-2}
(Q'-Q)\right\} \,.
\label{ratio}
\ee

Note that only the 2-- and 4--spin interactions contribute
when the action is written in the form of eq.~(\ref{14}).

In ref.~\cite{21} the $\Lambda$ parameter has been
calculated for a particular form of the 4--spin interactions.
Unfortunately we do not agree with the result quoted there.

Since in numerical simulations one has to work with a
restricted set of couplings, it is more informative
to have the $\Lambda$ parameter for that particular set.
For the couplings used in our simulations the ratio of
this $\Lambda$ parameter to that of the standard
lattice action is 8.17, as cited in eq.~(\ref{35}).

The $\Lambda$--parameter is rather sensitive to
the coefficients in the action.
 We did not attempt to calculate it for the whole set
of couplings in  $\A^*(\Sv)$, although this would be a
feasible task.
It is remarkable that while $\Lambda_{ST}$ is anomalously
small compared to those in a continuum regularization
(e.g. $\Lambda_{\overline{MS}}/ \Lambda_{ST}=27.21$),
$\Lambda_{FP}$ is much closer to its continuum counterparts.

\newpage

{\large\noindent\bf Figure captions}

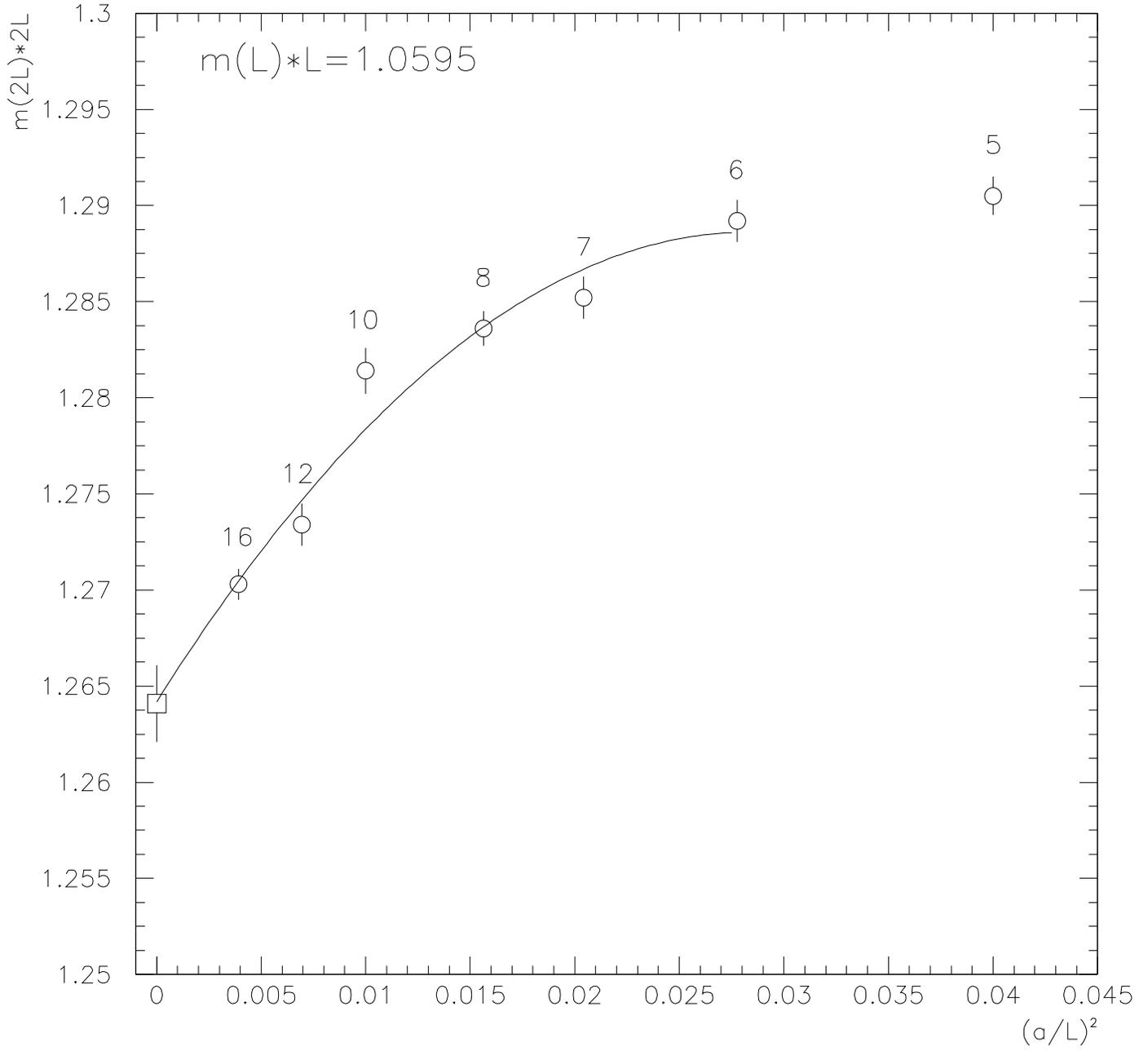
\noindent{Figure 1. The cut--off dependence of $m(2L)2L$ for
fixed value of $m(L)L=1.0595$, taken from \cite{2}.
The values of $L/a$ are indicated in the plot.
The curve is the result of a fit with a second order polynomial
in $(a/L)^2$. The open box shows the extrapolated value.}

\noindent{Figure 2. Flow of the couplings under RG transformation
 in the O($N$) non--linear $\sigma$--model.}

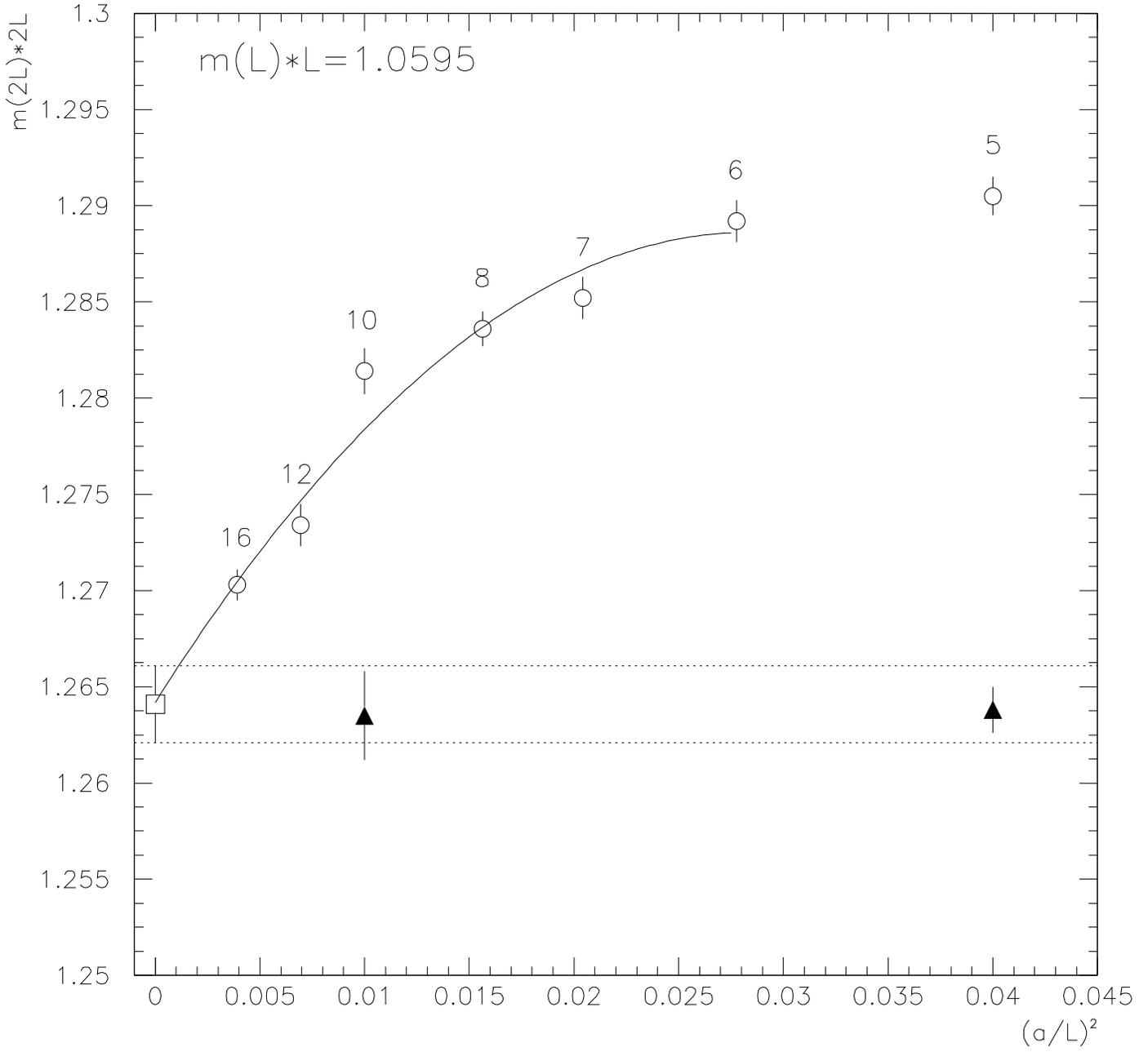
\noindent{Figure 3. The same as Fig.~1 with addition of two points
generated by using the FP--action. These data show no cut--off
dependence and agree with the value extrapolated to $a/L=0$ for
the standard action.}

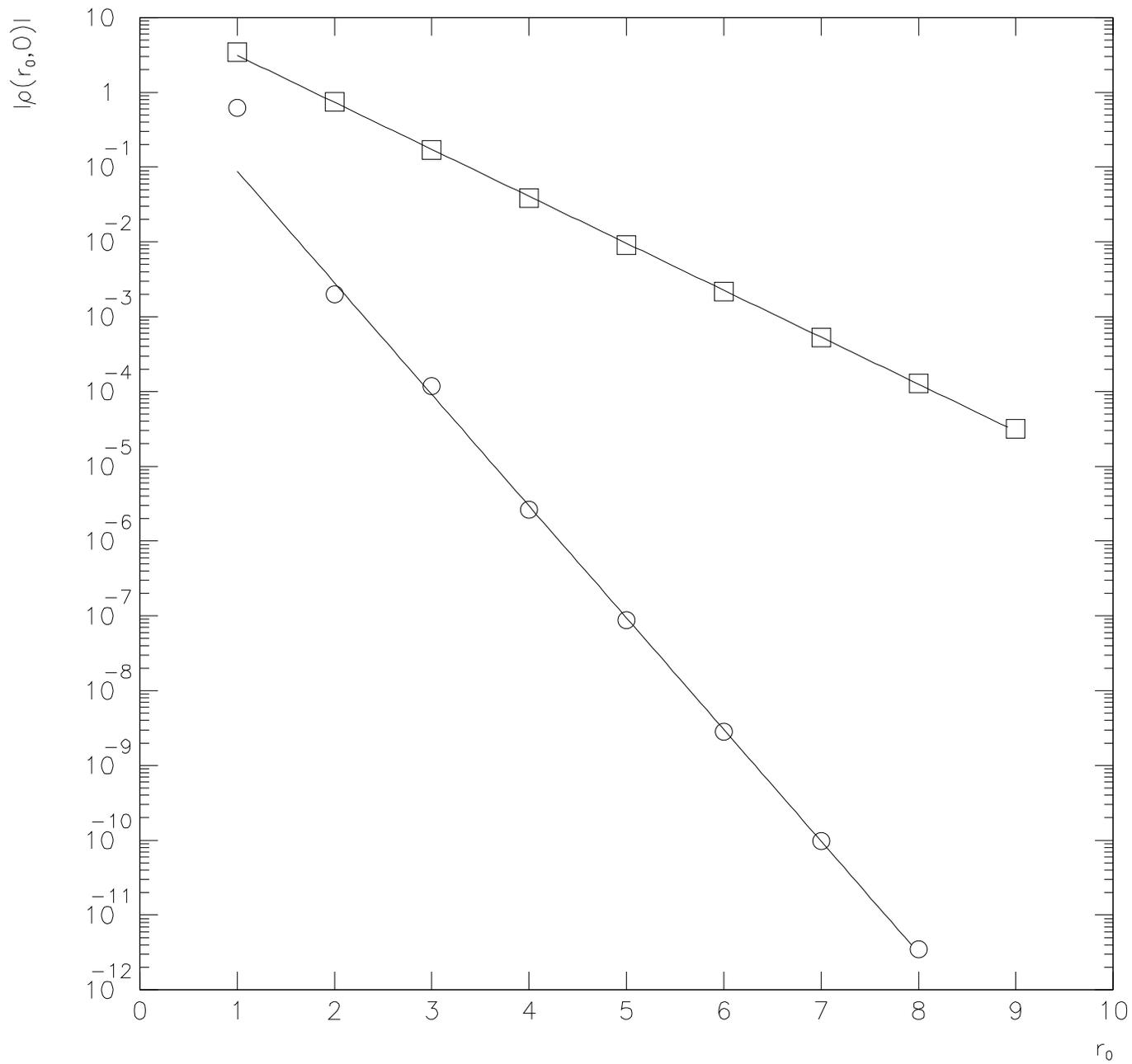
\noindent{Figure 4. Dependence of the two--spin interaction
coefficient $\rho(r)$ from the distance $r=(r_0,0)$ for
different choices of the block spin transformation.
The circles and boxes correspond to the optimal choice with
$\kappa=2$, and $\kappa=\infty$, respectively.}

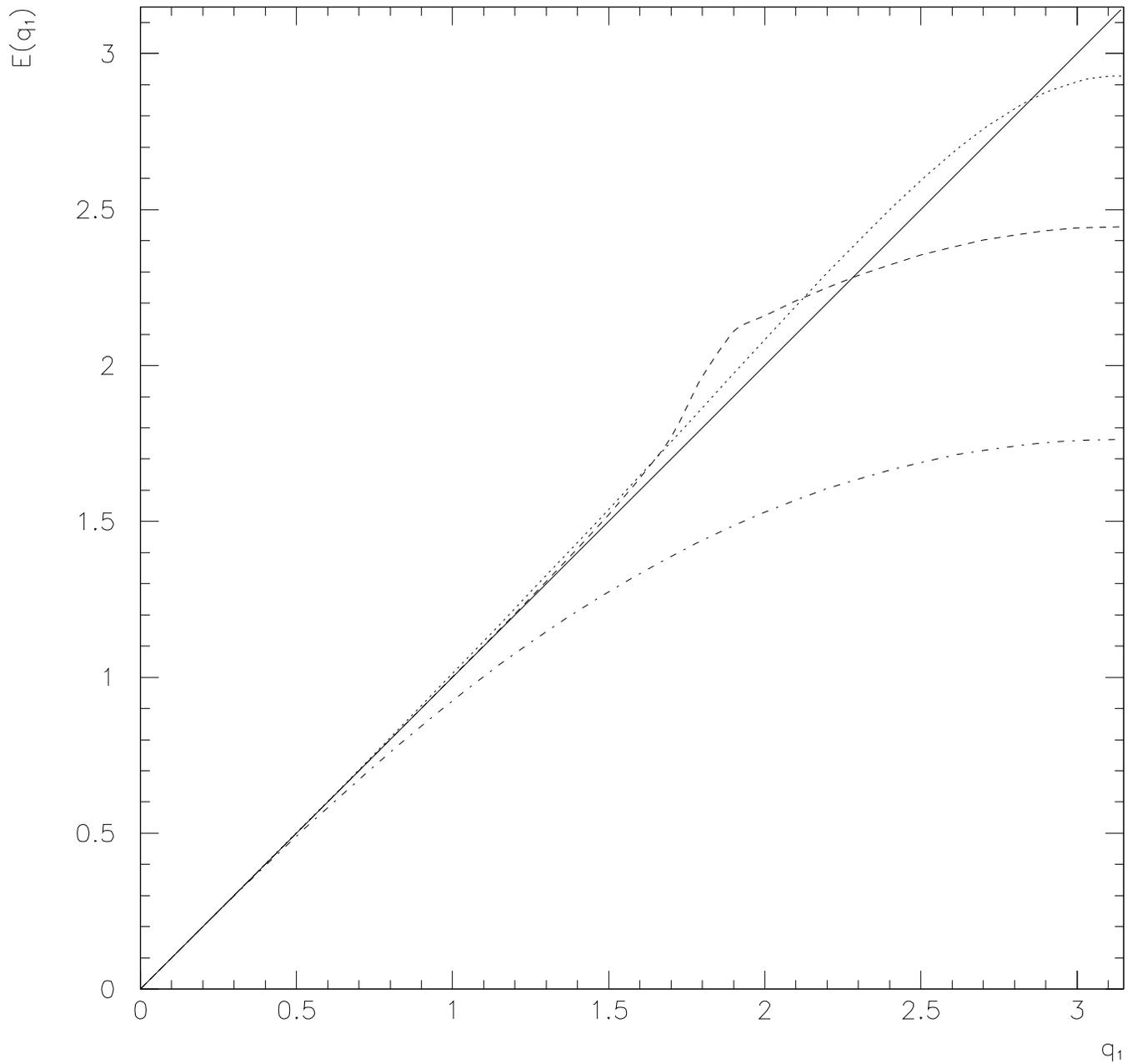
\noindent{Figure 5. The energy spectrum of the lattice regularized
free field theory for different lattice actions: the FP (exact)
(solid line), the approximation where the two leading couplings
of the FP are kept only (dotted line),
the standard action(dashed--dotted line)
and the Symanzik tree level improved action (dashed line).
Note that in the last case the eigenvalue becomes
complex above $|q_1|\approx 1.85$. For these values the real part is
plotted.}

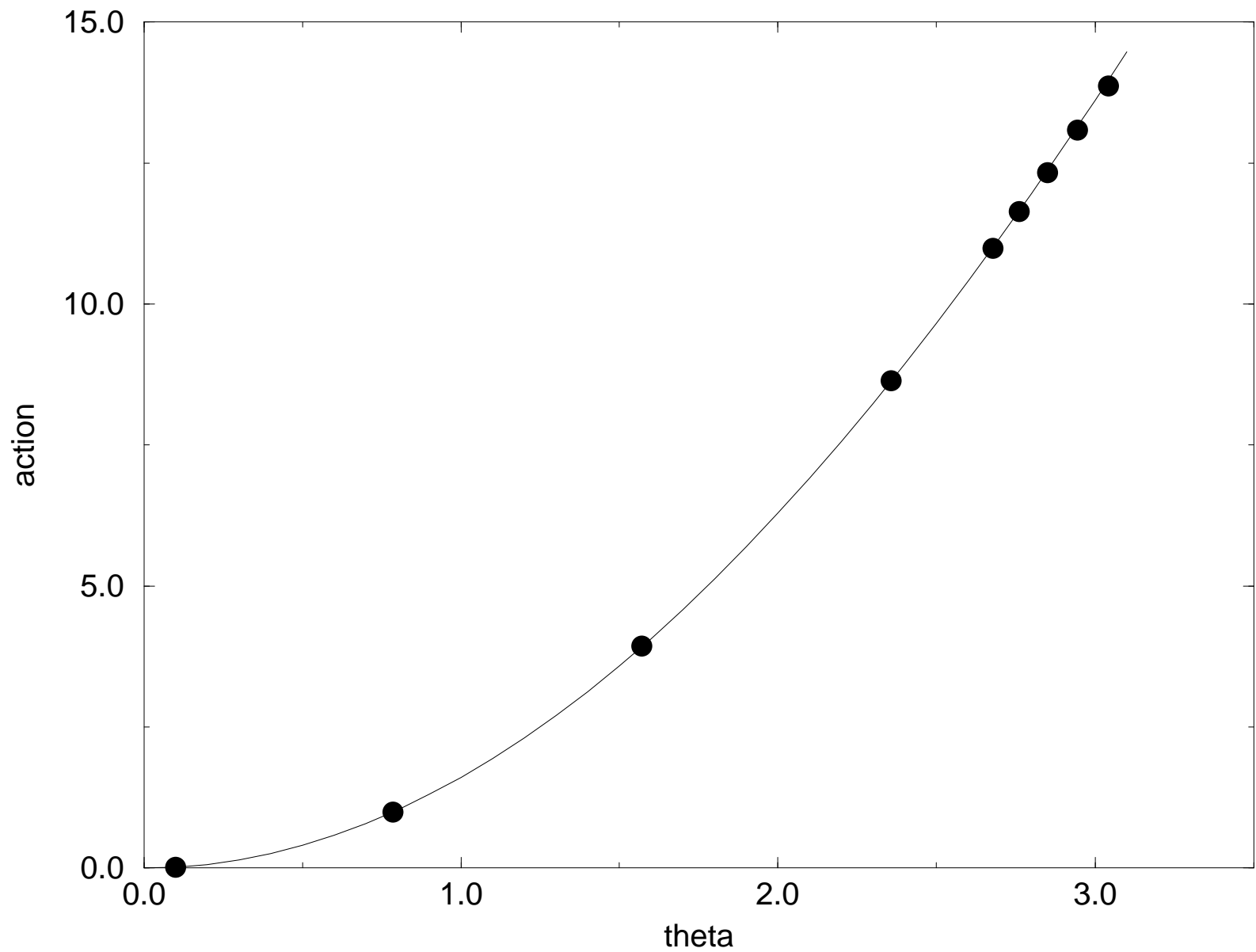
\noindent{Figure 6. The value of $\A^*(S)$ when one of the spins
is rotated by an angle $\vartheta$ with respect to a trivial
background of parallel spins. The curve is the contribution
from the quadratic and quartic terms in eq.~(12'),
$\A_1(\vartheta)=3.2403(\vartheta^2/2)-0.0476(\vartheta^2/2)^2$.}

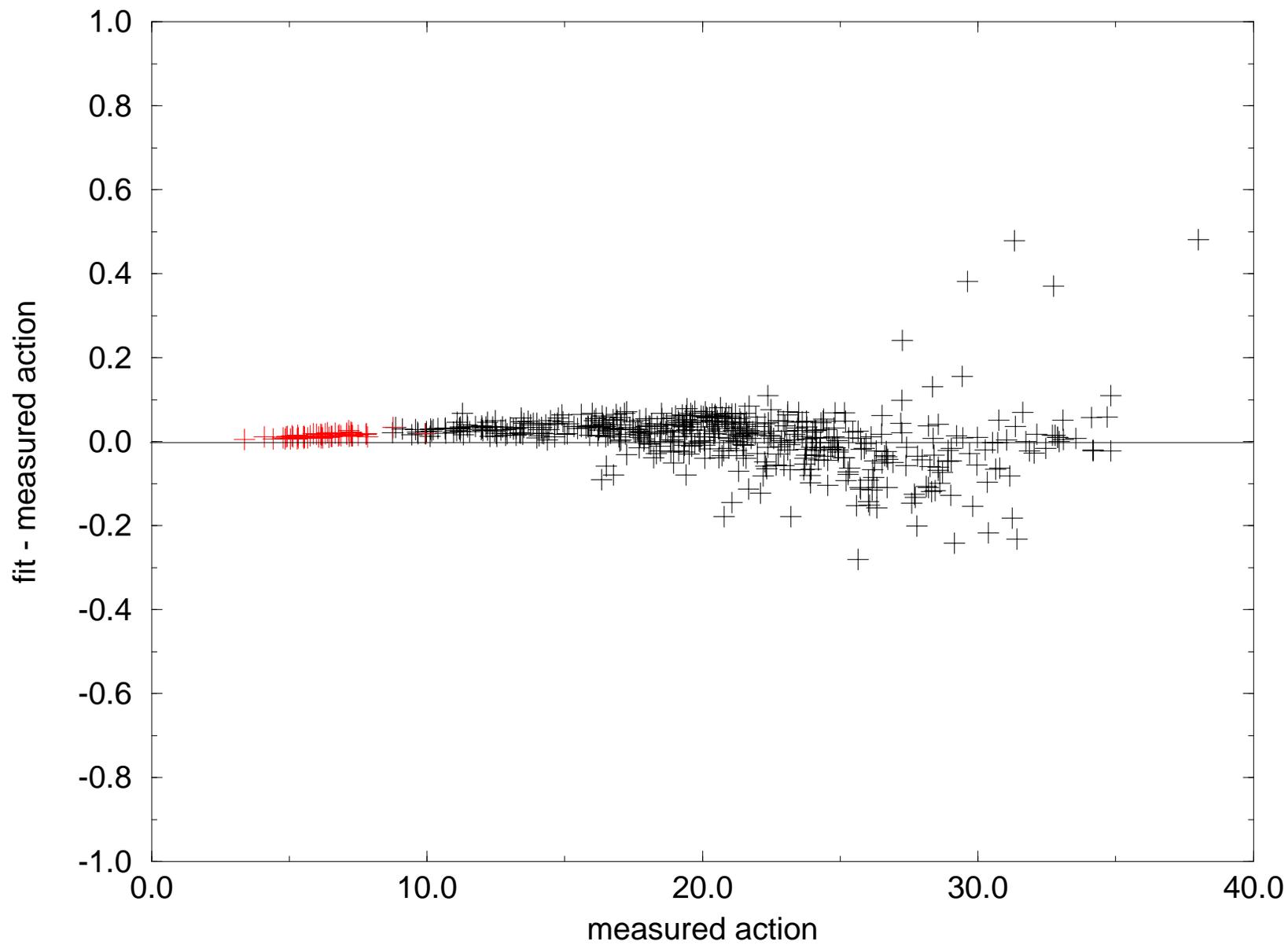
\noindent{Figure 7. The quality of the parametrization of $\A^*$
in Table~1 is shown in this scatter plot. The deviation from the
true value of the action is plotted against the value of the action
itself for $\approx 500$ configurations.}

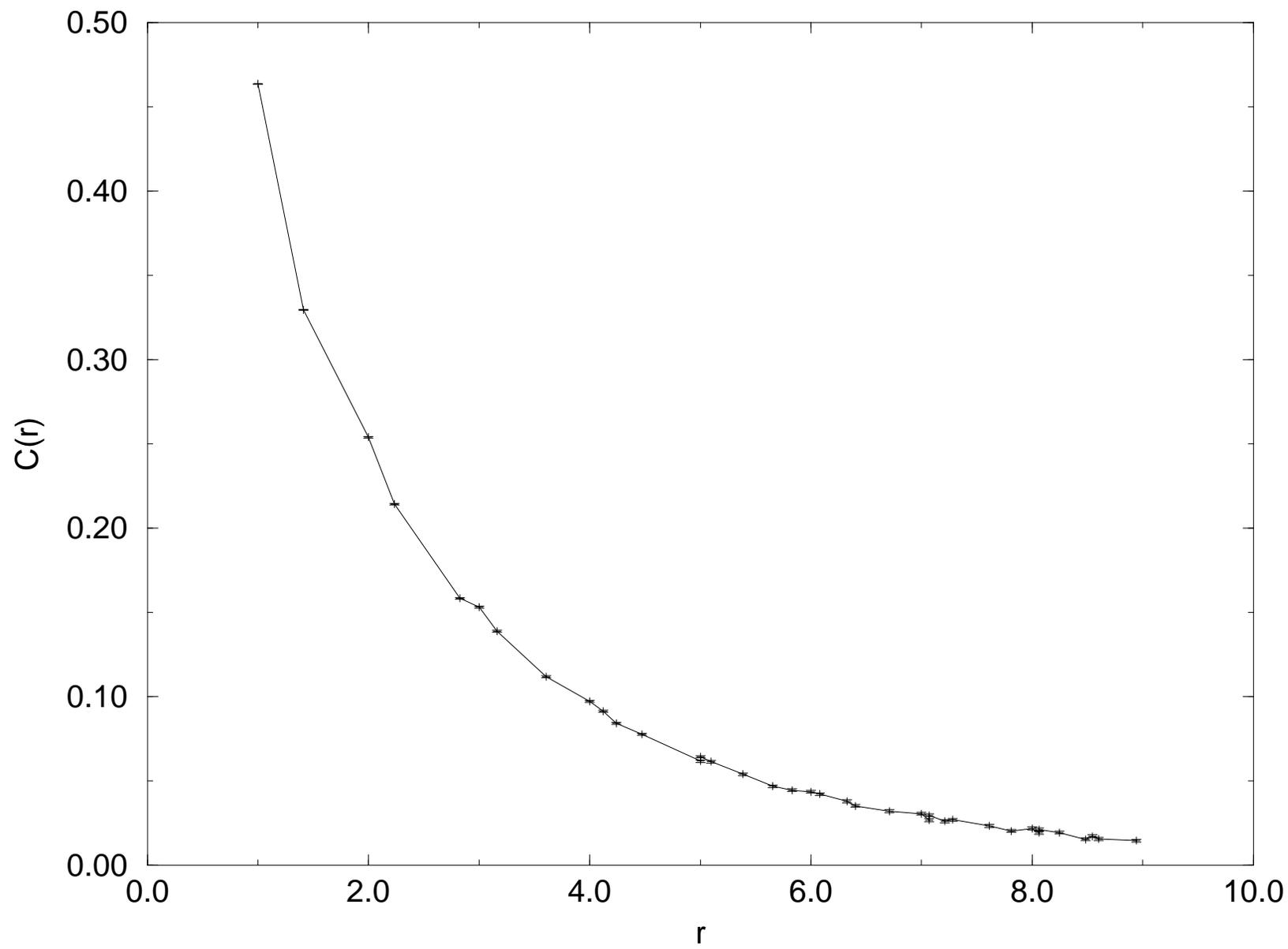
\noindent{Figure 8. The spin--spin correlation function
as measured with the standard action at $\beta_{ST}=1.18$
corresponding to a correlation length $\sim 3$. The points
are connected by piece--wise straight lines.}

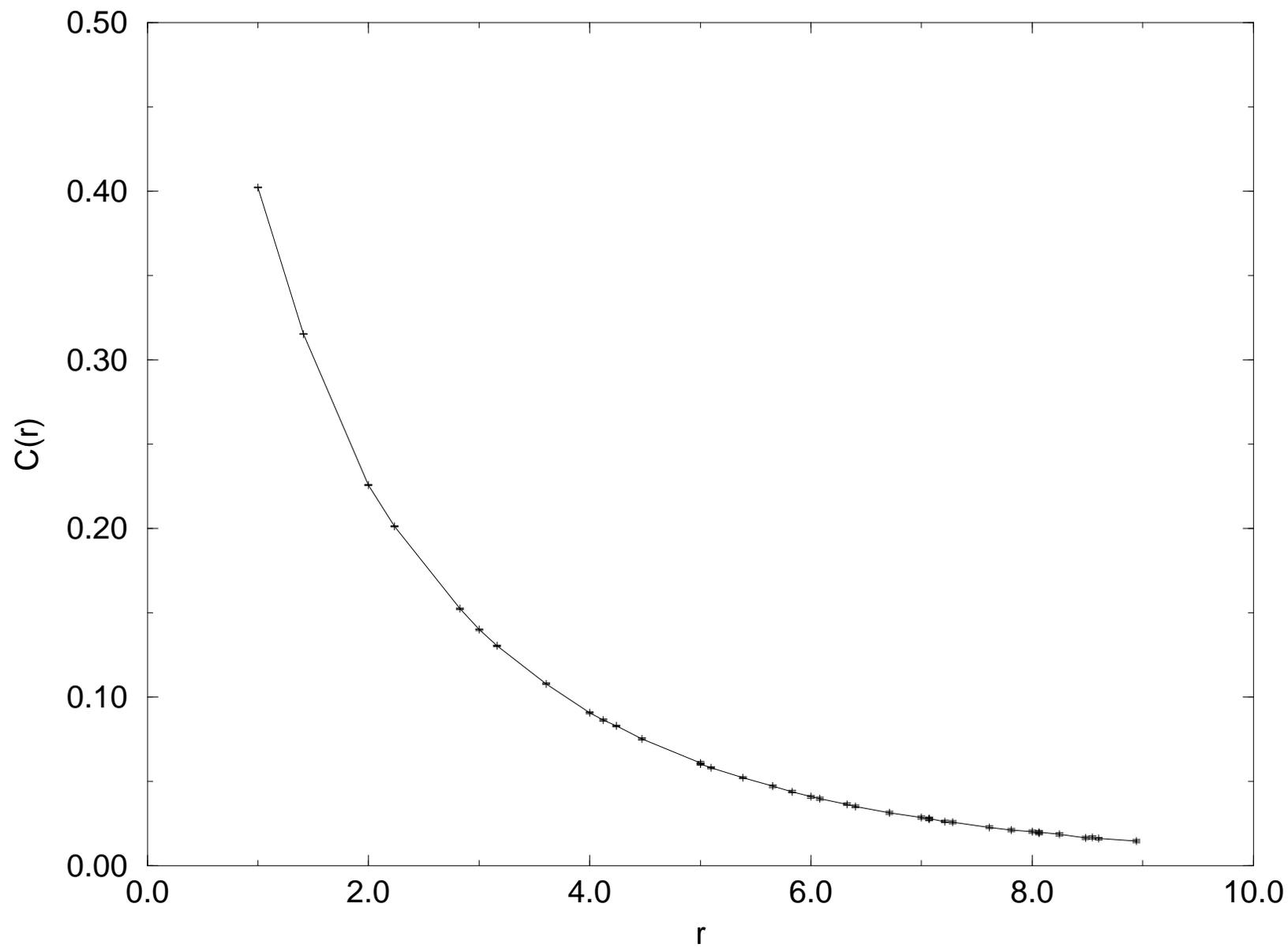
\noindent{Figure 9. The same as in Fig.~8 for the FP--action
at $\beta_{FP}=0.7$ corresponding to the same correlation length.}

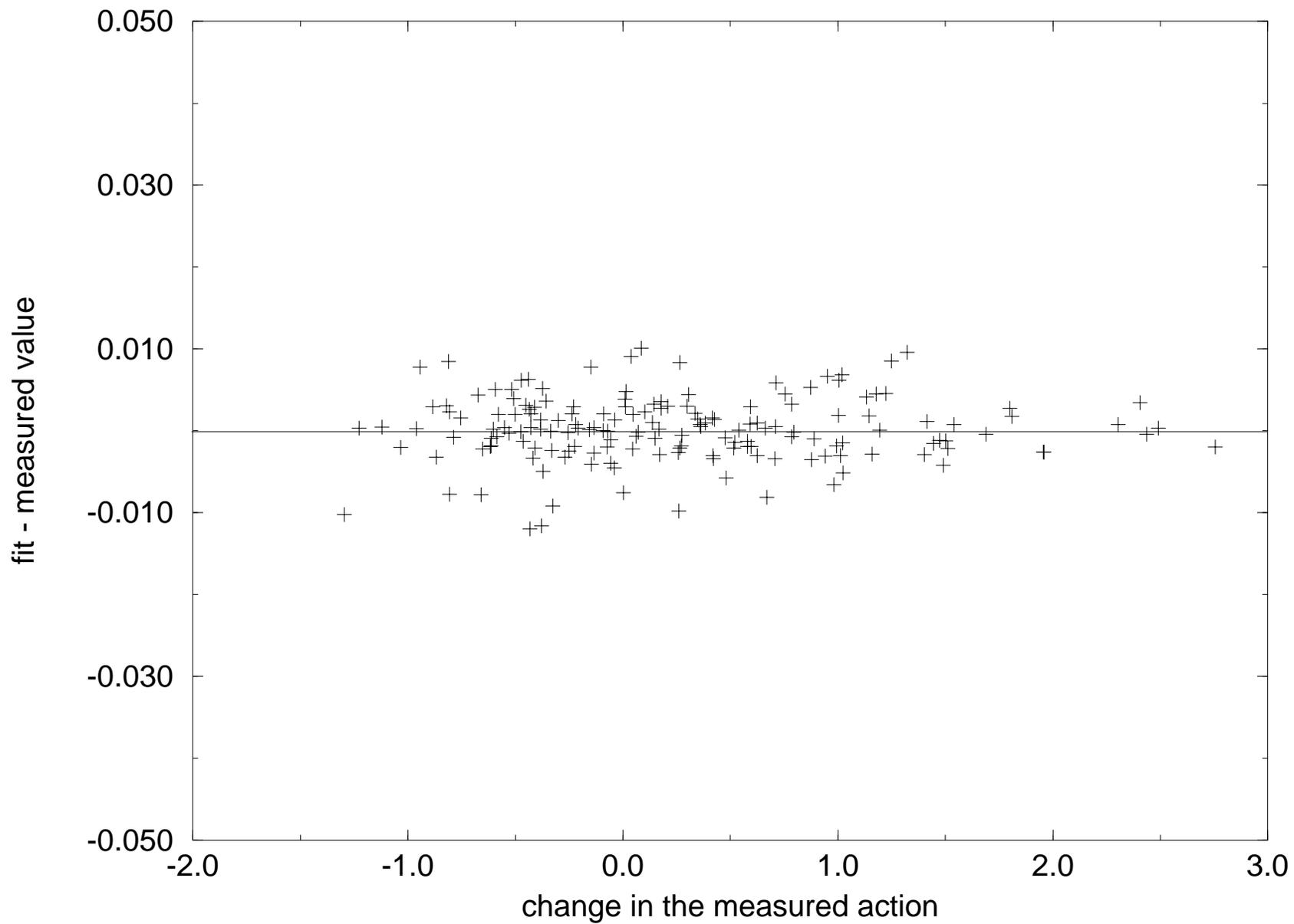
\noindent{Figure 10. The quality of the parametrization for
the action $\beta'\A'$ obtained after a block transformation on the
FP--action at $\beta_{FP}=1.0$ is shown here.
The fit included the interactions displayed in Table~4 plus
the (2,0), (2,1) and (2,2) couplings.}

\end{document}